\documentclass[%https://www.overleaf.com/project/5f51087b511da90001f73ed3
 reprint,
superscriptaddress,
 amsmath,
 amssymb,
 aps,
prb,
citeautoscript,
reprint,
floatfix
]{revtex4-2}

\usepackage{graphicx}% Include figure files
\usepackage{dcolumn}% Align table columns on decimal point
\usepackage{bm}% bold math
\usepackage{color}
\usepackage{amsmath}
\usepackage{url}
\usepackage[colorlinks=true,linkcolor=blue,citecolor=blue,urlcolor=blue]{hyperref}

\begin{document}

%\preprint{APS/123-QED}

\title{Effect of dynamical screening in the Bethe-Salpeter framework: Excitons in crystalline naphthalene }

\author{Xiao Zhang}
\altaffiliation[Now at ]{Department of Materials Science and Engineering, University of Michigan, Ann Arbor, MI 48105, USA.}
\affiliation{Department of Mechanical Science and Engineering, University of Illinois at Urbana-Champaign, Urbana, IL 61801, USA}
\author{Joshua A.\ Leveillee}
\altaffiliation[Now at ]{Oden Institute for Computational Engineering and Sciences, University of Texas, Austin, Texas, 78712, USA}
\affiliation{Department of Materials Science and Engineering, University of Illinois at Urbana-Champaign, Urbana, IL 61801, USA}
\author{Andr\'{e} Schleife}%
\email{schleife@illinois.edu}
\affiliation{Department of Materials Science and Engineering, University of Illinois at Urbana-Champaign, Urbana, IL 61801, USA}
\affiliation{Materials Research Laboratory, University of Illinois at Urbana-Champaign, Urbana, IL 61801, USA}
\affiliation{National Center for Supercomputing Applications, University of Illinois at Urbana-Champaign, Urbana, IL 61801, USA}

\date{\today}% It is always \today, today,

\begin{abstract}
Solving the Bethe-Salpeter equation (BSE) for the optical polarization functions is a first principles means to model optical properties of materials including excitonic effects.
One almost ubiquitously used approximation neglects the frequency dependence of the screened electron-hole interaction.
This is commonly justified by the large difference in magnitude of electronic plasma frequency and exciton binding energy.
We incorporated dynamical effects into the screening of the electron-hole interaction in the BSE using two different approximations as well as exact diagonalization of the exciton Hamiltonian.
We compare these approaches for a naphthalene organic crystal, for which the difference between exciton binding energy and plasma frequency is only about a factor of ten.
Our results show that in this case, corrections due to dynamical screening are about 15\,\% of the exciton binding energy.
We analyze the effect of screening dynamics on optical absorption across the visible spectral range and use our data to establish an \emph{effective} screening model as a computationally efficient approach to approximate dynamical effects in complex materials in the future.
\end{abstract}

\maketitle

\section{\label{sec:intro}Introduction}

Linear dielectric response is the underlying property that renders materials interesting for optoelectronic applications including solar cells, transistors, and displays, since  excitations of electrons control fundamental mechanisms of optical absorption and emission\cite{RevModPhys.77.1173,PHILLIPS196655,PhysRevLett.63.1719,rohlfing2000electron}.
Most state-of-the-art devices rely on traditional inorganic semiconductors that are well-studied from both experimental and theoretical perspective \cite{roundhill1999optoelectronic}.  
Apart from these, systems such as organic crystals have also been reported to have great potential e.g.\ as solar cells, sensors, transistors, and others \cite{liu1999optoelectronic,zimmerman2011mechanism,wang2018possibility,bredas2012conjugated,tang1986two}.
Singlet-triplet fission, for instance, can provide a novel mechanism that may enable design of more efficient, flexible solar cells \cite{wang2018possibility,zimmerman2011mechanism}. 
It is thus important to accurately model optical and excitonic properties for these materials, to make reliable predictions for potential applications and device design. 

Predictive first-principles simulations based on density functional theory (DFT) \cite{PhysRev.140.A1133, PhysRev.136.B864} and Fermi's golden rule have proven to be important in understanding optical absorption of many semiconductor materials \cite{rohlfing2000electron,PhysRevLett.63.1719,PhysRevB.45.11749,PhysRevB.25.6310}.
However, the lack of considering the electron-hole Coulomb interactions that dominate excited electronic states render traditional DFT-based techniques insufficient for describing optical absorption.
The independent-particle picture fails to provide accurate optical spectra and, in particular, does not provide information about excitonic effects that are critically important for applications, including organic solar cells \cite{dadsetani2015ab,Liu_2020,wang2016effect}.
To accurately model these, the \emph{screened} Coulomb interaction of excited electron-hole pairs needs to be considered and the Bethe-Salpeter equation (BSE) approach within many-body perturbation theory is commonly used \cite{rohlfing2000electron,bechstedt2016many}.
The solution of the BSE is a Green's function technique that allows to include excitonic effects in the first-principles description, leading to an accurate and commonly used theoretical-spectroscopy route to describe optical excitations.
It has proven successful in many studies that predict optical and excitonic properties of bulk semiconductors \cite{onida2002electronic,schleife2009optical,schleife2012ab,kang2019pushing,PhysRevB.92.045209}.

Within the BSE approach, the accurate description of dielectric screening is an important aspect of the underlying physics that is key to accurately simulating optical spectra.
While the screening of the electron-hole interaction is spatially inhomogeneous and dynamical in principle, especially the dynamical effects are not well explored in practice.
This is because the theoretical description of dynamical screening is challenging, the computational cost is high, and such effects are believed to be small in many traditional bulk semiconductors.
Hence, most of the BSE implementations currently used adopt a static, frequency-independent approximation of dielectric screening \cite{PhysRevB.92.045209,DESLIPPE20121269,Sangalli_2019,Vorwerk_2019,BLUM20092175,Giantomassi2011}.
This approximation neglects the rearrangement of the electrons upon forming electron-hole pairs, i.e.\ the dynamical evolution of the screening \cite{bechstedt2016many}.

Whether \emph{electronic} screening dynamics can be neglected, however, is related to the relative ratio of the plasma frequency and the exciton binding energy of a material \cite{bechstedt2016many,rohlfing2000electron}.
In particular, electronic dynamical effects cannot be neglected when exciton-binding energies are comparable to the plasma frequency.
Examples of large exciton binding energies on the order of few hundreds of meV to a few eV include low-dimensional materials \cite{qiu2016screening,gao2016dynamical,zhu2015exciton,ugeda2014giant} and organic crystals \cite{hummer2005oligoacene,dadsetani2015ab,Liu_2020,wang2016effect}. 
Consequently, there are indeed computational studies of organic crystals and doped 2D materials\cite{leng2015excitons,gao2016dynamical} that report large corrections on the order of a few hundred meV for the exciton binding energy due to electronic screening dynamics.
However, different approximations of incorporating dynamical screening effects, treating the dynamical screening as a first-order perturbation\cite{rohlfing2000electron} or using an effective static screening function\cite{gao2016dynamical} have not been systematically compared before to each other or to exactly solving the dynamical BSE. 

In this work, we discuss different approximations for incorporating dynamical screening into the solution of the BSE. 
The challenges are at least two-fold:
Dynamical screening of the electron-hole interaction complicates the many-body perturbation theory framework, since the resulting BSE depends on two frequencies, preventing a closed-form equation for a single-frequency dependent polarization function \cite{bechstedt2016many,rohlfing2000electron,blase2018bethe}.
While this can be overcome using the Shindo approximation \cite{shindo1970effective,bechstedt2016many},
the resulting BSE eigenvalue problem parametrically depends on the frequency and requires sampling of many frequency points, significantly increasing the computational cost.
Rohlfing and Louie proposed a perturbative treatment and used it to examine dynamical screening in molecular SiH$_4$\cite{rohlfing2000electron}. 
This approach is also used to examine dynamical effects in biological organic materials such as photoactive yellow protein and dicyanovinyl-substituted oligothiophenes \cite{ma2009excited,baumeier2012excited,ma2009modeling}. 
While it provides an efficient way to incorporate dynamical screening for a small number of excitonic states e.g.\ around the absorption edge, it is not directly applicable for simulations of optical spectra, where a large number of excitonic states across a certain energy range is required.
In addition, this approach approximates the true excitonic wave function by the static one, which is only valid when dynamical effects are small.

In this work we follow Refs.\ \onlinecite{bechstedt2016many, hybertsen1986electron} in using the Shindo approximation and a plasmon pole model for the analytical integration of the frequency-dependent dielectric function
to obtain an expression for a single frequency-dependent dynamical BSE.
We then implement and compare different approximations to numerically solve the dynamical BSE, including a static model with effective screening \cite{gao2016dynamical}, the above-mentioned first-order perturbation approach \cite{rohlfing2000electron}, and exact diagonalization of the Hamiltonian.
By solving the dynamical BSE directly on a frequency grid, we were able to examine not only the effect of dynamical screening on exciton-binding energies, but also on optical spectra.
Our results show that while approximate treatments provide reasonable estimates of the magnitude of spectral shifts due to screening dynamics close to the absorption onset, small qualitative differences remain compared to the exact solution for excitonic states higher in energy.
In addition, we show that an \emph{effective} static screening, derived within the dynamical screening framework \cite{gao2016dynamical,PhysRevLett.127.067401}, requires only the lowest exciton-binding energy as input and still provides a good description of spectra.
It provides a computationally tractable alternative e.g.\ for studying complex or large numbers of materials.

In this work we use crystalline naphthalene as an example.
For this material, previous theoretical calculations report exciton binding energies of 0.9 eV, underestimating experimental measurements of 1.0\,--\,1.5 eV\cite{hummer2005oligoacene}.
Since this exciton binding energy is about 10\,\% of the plasma frequency, dynamical electronic screening can become important \cite{rohlfing2000electron, bechstedt2016many, gao2016dynamical,ma2009excited}.
Our work provides a quantitative understanding of the importance of dynamical electronic screening and provides guidance for appropriate regimes of using different approximations when studying optical and excitonic properties of more complicated materials.

\section{\label{sec:theory}Theoretical approach}

The theoretical description of excitonic effects in this work starts from the Bethe-Salpeter equation (BSE) for the macroscopic ($M$) optical polarization function $P^{M}$.
It follows from Hedin's equations for interacting electrons \cite{Hedin:1965} and describes the probability amplitude of the process of annihilating an electron at ($\mathbf{r}'_2,t'_2$) after creating one at ($\mathbf{r}'_1,t'_1$), together with annihilating a hole at ${\left(\mathbf{r_2}, t_2\right)}$ after creating one at ${\left(\mathbf{r_1}, t_1\right)}$.
In  reciprocal and frequency space the full BSE reads\cite{bechstedt2016many}
\begin{widetext}
\begin{equation}
\label{eq:fullbse}
\begin{split}
P^M(\lambda_1\lambda'_1,\lambda_2\lambda'_2,z_nz_m)=&-i\hbar G_{\lambda_1}(z_n)G_{\lambda'_1}(z_n-z_m)\times\{\delta_{\lambda_1\lambda'_2}\delta_{\lambda'_1\lambda_2}\\
&+\frac{1}{-i\hbar\beta}\sum_{n'}\sum_{\lambda_3\lambda_4}
\left[-W^{\lambda_1\lambda_3}_{\lambda'_1\lambda_4}(z_n-z_{n'})+2\bar{v}^{\lambda_1\lambda'_1}_{\lambda_3\lambda_4}\right]\times P^M(\lambda_3\lambda_4,\lambda_2\lambda'_2,z_{n'}z_m).\}
\end{split}
\end{equation}
\end{widetext}
Here, $\beta=1/(k_\mathrm{B} T)$ where $k_\mathrm{B}$ is the Boltzmann constant and $T$ is temperature.
$z_n$ and $z_{n'}$ are Fermionic Matsubara frequencies, corresponding to the Fourier components of the time difference between $t_1$ and $t'_1$, $z_m$ is the Bosonic Matsubara frequency, corresponding to the Fourier component of the time difference between $t_1$ and $t_2$.
$\lambda$ are indices for all single-particle electronic states.
$G_{\lambda}(z)=1/(\hbar z-E_{\lambda})$ are single-particle Green's functions\cite{PhysRevB.29.5718, rohlfing2000electron} with $E_{\lambda}$ being the energy of the single-particle electron and hole state $\lambda$.
$W$ and $\bar{v}$ are the screened and the short-range bare Coulomb interaction of electrons and holes, respectively.
It can be seen that the polarization function in Eq.\ \eqref{eq:fullbse} depends on \emph{two} frequency arguments, $z_n$ and $z_m$. 

To describe optical excitation due to absorption of a single photon, one needs to obtain the polarization function that depends on only \emph{one} frequency.
In principle, this can be obtained by summing Eq.\ \eqref{eq:fullbse} over $n$, \cite{bechstedt2016many}
\begin{equation}
\label{eq:polarization}
P^M(\lambda_1\lambda'_1,\lambda_2\lambda'_2,z_m)=\frac{1}{-i\hbar\beta}\sum_nP^M(\lambda_1\lambda'_1,\lambda_2\lambda'_2,z_nz_m).
\end{equation}
In practice, evaluating Eq.\ \eqref{eq:polarization} is difficult for two reasons:
First, to obtain each polarization function on the right-hand side, a complicated matrix problem needs to be solved that involves the $n'$-sum over the frequency-dependent screened Coulomb interaction in Eq.\ \eqref{eq:fullbse}.
Second, this procedure needs to be done many times for different $z_n$ in order to evaluate the sum in Eq.\ \eqref{eq:polarization}. 

\subsection{Static Bethe-Salpeter equation}

The standard approach to avoiding these difficulties is to neglect the frequency dependence of the screening, i.e.\ assuming
\begin{equation}
\label{eq:static}
W(z_n-z_{n'})\equiv W(0),
\end{equation}
where $W(0)$ is the zero-frequency, static limit.
It can be shown that with this approximation one can insert Eq.\ \eqref{eq:fullbse} into Eq.\ \eqref{eq:polarization} and obtain a problem that only involves the polarization function that depends on a single frequency argument\cite{bechstedt2016many,PhysRevB.29.5718,rohlfing2000electron}
\begin{widetext}
\begin{equation}
\label{eq:bse_static}
P^M(\lambda_1\lambda'_1,\lambda_2\lambda'_2,z_m)=\frac{f(\lambda_1)-f(\lambda_1')}{E_{\lambda_1}-E_{\lambda_1'}-\hbar z_m}\times\{\delta_{\lambda_1\lambda'_2}\delta_{\lambda'_1\lambda_2}+\sum_{\lambda_3\lambda_4}
\left[-W^{\lambda_1\lambda_3}_{\lambda'_1\lambda_4}+2\bar{v}^{\lambda_1\lambda'_1}_{\lambda_3\lambda_4}\right]\times P^M(\lambda_3\lambda_4,\lambda_2\lambda'_2,z_m).\}
\end{equation}
\end{widetext}
The Green's functions in Eq.\ \eqref{eq:fullbse} result in the term $\frac{f(\lambda_1)-f(\lambda_1')}{E_{\lambda_1}-E_{\lambda_1'}-\hbar z_m}$, where $f(\lambda)$ is the occupation factor of state $\lambda$.
The crucial difference to Eq.\ \eqref{eq:fullbse} is that Eq.\ \eqref{eq:bse_static} contains only one frequency argument $z_m$ and the complicated sum over $n$ in Eq.\ \eqref{eq:polarization} is avoided.
Subsequently, Eq.\ \eqref{eq:bse_static} is transformed into a generalized eigenvalue problem\cite{fuchs2008efficient}.

From now on, we consider translational invariance, fully occupied valence states, and entirely empty conduction states, as is the case in semiconductor crystals at low temperature.
This turns $\lambda \rightarrow c\mathbf{k}, v\mathbf{k}$, and the standard BSE Hamiltonian is obtained\cite{rohlfing2000electron,bechstedt2016many,fuchs2008efficient} as
\begin{equation}
\label{eq:bse}
\begin{split}
\hat{H}_{vc\mathbf{k},v'c'\mathbf{k'}}=&(E_{c\mathbf{k}}-E_{v\mathbf{k}})\delta_{vv'}\delta_{cc'}\delta_{\mathbf{kk'}}\\&+2\bar{v}^{v'c'\mathbf{k}'}_{vc\mathbf{k}}-W^{v'c'\mathbf{k}'}_{vc\mathbf{k}},
\end{split}
\end{equation}
where $E_{c\mathbf{k}}$ and $E_{v\mathbf{k}}$ are the energies of the electronic state at point $\mathbf{k}$ in reciprocal space, and $c$ and $v$ represent conduction and valence band index, respectively. 
The term $\bar{v}^{v'c'\mathbf{k}'}_{vc\mathbf{k}}$ describes the bare Coulomb interaction, which is a short-range exchange term, and $W^{v'c'\mathbf{k}'}_{vc\mathbf{k}}$ describes the screened electron-hole Coulomb interaction that in the static approximation is computed using the inverse $\mathbf{q}$-dependent dielectric matrix $\varepsilon^{-1}(\mathbf{q}, \omega=0)$.
Solving the eigenvalue problem for the Hamiltonian in Eq.\ \eqref{eq:bse} provides pair resonance energies $E_\Lambda$ and eigenfunctions $\mathbf{A}_\Lambda$ for excitonic states indexed by $\Lambda$.
These are used to compute the dielectric function that can be compared to experiment, and to analyze exciton binding energies. 
The connection of the polarization function $[P(\omega)]$ from the solution of the eigenvalue problem, and the dielectric function, is identical in the static and dynamical screening case, $[\varepsilon(\omega)=1-vP(\omega)]$, see Ref. 18 for details.
\vspace{0.3 cm}
\subsection{\label{sec:dynbse}Dynamical Bethe-Salpeter equation}

To preserve the frequency dependence of $W$, an alternative way of obtaining a single-frequency dependent polarization function is through Shindo's approximation\cite{shindo1970effective}.
Instead of Eq.\ \eqref{eq:static}, this approximation expresses the two-frequency dependent polarization function $P^{M}(z_nz_m)$ in Eq.\ \eqref{eq:fullbse} directly in terms of the Green's function of non-interacting electrons and holes and the one-frequency dependent polarization function $P^{M}(z_m)$, see Eq. (S1) in the supplemental information\cite{supplement}.
This approximation leads to an expression for the single-frequency dependent polarization function that takes a very similar form as Eq.\ \eqref{eq:bse_static}, with the frequency-independent screened Coulomb interaction $W$ replaced by an effective, frequency-dependent $\tilde{W}(z_m)$ \cite{bechstedt2016many} (see Eq. (S3) in the supplemental information\cite{supplement}).
Considering only real frequencies involved in optical excitations ($z_m\rightarrow\omega$) allows the transformation into an eigenvalue problem for the frequency-dependent BSE Hamiltonian\cite{bechstedt2016many}
\begin{equation}
\label{eq:hamiltonian_dyn}
\begin{split}
\tilde{H}_{vc\mathbf{k},v'c'\mathbf{k'}}(\omega)=&(E_{c\mathbf{k}}-E_{v\mathbf{k}})\delta_{vv'}\delta_{cc'}\delta_{\mathbf{kk'}}\\&+2\bar{v}^{v'c'\mathbf{k}'}_{vc\mathbf{k}}-\tilde{W}^{v'c'\mathbf{k}'}_{vc\mathbf{k}}(\omega).
\end{split}
\end{equation}
Compared to Eq.\ \eqref{eq:bse}, the frequency dependence of Eq.\ \eqref{eq:hamiltonian_dyn} comes from the effective, frequency-dependent screened Coulomb interaction. 
The effective frequency-dependent $\tilde{W}(\omega)$ takes the form\cite{rohlfing2000electron,bechstedt2016many}
\begin{widetext}
\begin{equation}
\label{equ:w}
\begin{split}
\tilde{W}^{v'c'\mathbf{k}}_{vc\mathbf{k}}(\omega)=&\frac{1}{V}
\sum_{\mathbf{q, G, G'}}v\left(\sqrt{|\mathbf{q}+\mathbf{G}||\mathbf{q}+\mathbf{G'}|}\right)B_{cc',\mathbf{kk}'}(\mathbf{q}+\mathbf{G})B_{vv',\mathbf{kk}'}^{*}(\mathbf{q}+\mathbf{G'}) \\
&\times \Big\{\delta_{\mathbf{GG'}}+\int_0^{\infty}\frac{d\hbar \omega '}{\pi}\text{Im}\varepsilon^{-1}(\mathbf{q}+\mathbf{G},\mathbf{q}+\mathbf{G'},\omega ')\times \\
&\times\left[\frac{1}{\hbar\omega '+E_{c\mathbf{k}}-E_{v'\mathbf{k}'}-\hbar \omega}+\frac{1}{\hbar\omega '+E_{c'\mathbf{k}'}-E_{v\mathbf{k}}-\hbar \omega}\right]\Big\}\delta_{\mathbf{q},\mathbf{k-k'}},
\end{split}
\end{equation}
\end{widetext}
where $v$ is the Coulomb potential in reciprocal space.
In addition, $V$ is the volume of the unit cell, $\mathbf{q}$ represent the reciprocal wavevector, $\mathbf{G}$ and $\mathbf{G}'$ are are reciprocal lattice vectors. 
$\varepsilon(\mathbf{q+G, q+G}', \omega ')$ is the frequency and wavevector dependent dielectric function.  
The terms $B_{cc',\mathbf{kk'}}$ and $B_{vv',\mathbf{kk'}}$ are the Bloch integrals that account for the overlap between single particle Bloch wave functions\cite{bechstedt2016many}: 
\begin{equation}
B_{cc',\mathbf{kk'}}(\mathbf{q+G})=\delta_{\mathbf{q, k-k'}}\langle u_{c\mathbf{k}}|e^{i\mathbf{G}\cdot \mathbf{r}}|u_{c'\mathbf{k}'}\rangle, 
\end{equation}
where $u_{c\mathbf{k}}(r)$ represent the Bloch-periodic part of the Kohn-Sham state $\psi_{c\mathbf{k}}$. The frequency integral in the second term inside the curly brackets results from Shindo's approximation\cite{bechstedt2016many}, since the two-frequency dependence is replaced by a sum over single-frequency dependent polarization functions\cite{bechstedt2016many,shindo1970effective}. 
We note that the Shindo approximation uses the screened Coulomb interaction $W$, that is also part of the GW approximation, and turns it into $\tilde{W}$ via Eq. (S4) of the supplemental material. $\tilde{W}$ is more complicated due to the additional energy denominator terms.

We refer to the supplemental information Eq. (S3) for the single frequency dependent polarization function using Shindo's approximation\cite{supplement} and Ref.\ \onlinecite{bechstedt2016many} for the derivation of Eq.\ \eqref{equ:w}, as well as for complete details on how to obtain the eigenvalue problem from the BSE, which is identical in the static and dynamic case.
The form of the frequency dependent screened Coulomb potential Eq.\ \eqref{equ:w} has also been derived in multiple other references \cite{shindo1970effective,rohlfing2000electron,bechstedt1980theory}.
Shindo's approximation is argued to be a first-order approximation with respect to the dynamical nature of the screened potential\cite{bechstedt2016many,scharf2019dynamical,bornath1999two,zimmermann1971,zimmermann1978}.
Without assuming static screening in the BSE, the Shindo approximation, Eq. (S1), requires a small difference of the dynamic potential and an effective static potential. It uses only the zero’th order of an expansion in the difference of these two potentials, leading to Eq. (S3) and Eq. (S4). 
We show later in Sec. \ref{sec:dyn} and \ref{sec:omegap} that in our work this requirement is fulfilled since the dependence of the excitation energy on frequency is an order of magnitude smaller than the excitation energy itself. We will show that the dynamical correction does not exceed 15\% of the exciton binding energy.
However, studying its validity quantitatively is very hard and has not been accomplished so far.
In this work, we analyze dynamical screening effects within the framework of Shindo's approximation, and do not consider any effects beyond.
With Eqs.\ \eqref{eq:hamiltonian_dyn} and \eqref{equ:w}, two aspects need to be addressed to solve the dynamical problem.
First, an eigenvalue problem needs to be solved similar to the static case, however, now with a frequency-dependent BSE Hamiltonian.
Second, one needs to evaluate the frequency-dependent screened Coulomb interaction, Eq.\ \eqref{equ:w}. 
In the following, we discuss practical ways to address the first aspect in Sec.\ \ref{sec:dyn_eig}, and the second aspect in Sec.\ \ref{sec:dynscreen}. 

\subsection{\label{sec:dyn_eig}Dynamical eigenvalue problem}

A dynamical eigenvalue problem needs to be solved for the frequency-dependent BSE Hamiltonian Eq.\ \eqref{eq:hamiltonian_dyn},
\begin{equation}
\label{eq:bsefreq}
\tilde{H}(\omega)\mathbf{A}_{\Lambda}(\omega)=E_{\Lambda}(\omega)\mathbf{A}_{\Lambda}(\omega),
\end{equation}
to obtain the  frequency-dependent excitonic eigenvalues and eigenfunctions.
Different from the static case, where this set of solutions directly provides excitation energies, in the case of dynamical screening, one needs to find the solution of \cite{bechstedt2016many}
\begin{equation}
\label{eq:solution}
 E_\Lambda(\omega)=\hbar\omega.
\end{equation}
Physically, this represents the condition where the energy of excitonic state $\Lambda$ equals the energy of the absorbed photon and it amounts to identifying the state $\Lambda$ that was computed using the corresponding photon frequency.
In the following, we discuss three different approaches to accomplish this:
Exact diagonalization of the dynamical Hamiltonian, a perturbative treatment of the problem \cite{rohlfing2000electron}, and an \emph{effective} static screening approximation\cite{gao2016dynamical,PhysRevLett.127.067401}.

In the exact diagonalization approach, the excitation energy $E_\Lambda(\omega)$ can be obtained by sampling the frequency $\omega$ on a grid and solving one eigenvalue problem at each frequency point. 
Subsequently, Eq.\ \eqref{eq:solution} can be solved via interpolation of this data or using the nearest data point on the frequency grid that minimizes $E_{\Lambda}(\omega)-\hbar\omega$. 
Compared to the static BSE, this increases the complexity by at least a factor of $N$, where $N$ is the number of frequency sampling points.
We note that this computational cost can be somewhat mitigated using efficient solvers of eigenvalue problems, such as the ChASE library \cite{chase22}, that we recently interfaced with our BSE code \cite{chase21}, demonstrating speedups on the order of a factor of five in solving the static BSE.

The perturbative approach to solving Eq.\ \eqref{eq:solution} was proposed by Rohlfing and Louie \cite{rohlfing2000electron}.
It treats the dynamical effect of the screened Coulomb potential as a first-order perturbation to the solutions of the static BSE.
The solutions $E^{\text{sta}}_{\Lambda}$ of the static eigenvalue problem for each excitonic state $\Lambda$ are used as the input frequency $\hbar\omega$ in Eq.\ \eqref{equ:w} to the dynamical screening function $\tilde{W}(\omega)$.
Next, the difference between the resulting approximated dynamical screening potential and the static screening potential $\tilde{W}(E^\mathrm{sta}_\Lambda)-W^{\text{sta}}$ is treated as a first-order perturbation, so that the solution for each state $\Lambda$ becomes
\begin{equation}
\label{eq:perturb}
E^{\text{dyn}}_{\Lambda}\approx E^\mathrm{sta}_{\Lambda}+\langle A_\Lambda|\tilde{W}(E^\mathrm{sta}_\Lambda)-W^\mathrm{sta}|A_{\Lambda}\rangle,
\end{equation}
where $|A_\Lambda\rangle$ are the eigenfunctions of the static BSE Hamiltonian Eq.\ \eqref{eq:bse}.

Validity of the perturbative treatment requires two conditions:
First, that $E^{\text{sta}}_\Lambda$ is reasonably close to the true solution such that evaluating $\tilde{W}(\omega)$ at $\hbar\omega=E^{\text{sta}}_\Lambda$ is close to the true dynamical screening function for each state $\Lambda$, and second, that the difference $\tilde{W}(E^\mathrm{sta}_\Lambda)-W^\mathrm{sta}$ is small, so that dynamical effects can be considered as a first-order perturbation. 
Ref.\ \onlinecite{rohlfing2000electron} recommends to iterate several times and reports quick convergence, and indeed we verified that the solution converges within two to three steps.
Instead of solving the entire problem on a frequency grid, this approach focuses on a few specific excitonic states and only updates the energy of those states based on the static solutions, leaving the excitonic wave functions unchanged. 
This provides a fast route to solving the dynamical problem especially when only a few excitonic states, e.g.\ the lowest ones, are of interest.
However, its validity needs to be examined, in particular for systems where dynamical effects are significant, since in this case, the solutions obtained through the static approximation can differ significantly from the dynamic ones. 

As can be seen in Eq.\ \eqref{eq:perturb}, the perturbative approach requires evaluating the screened interaction $\tilde{W}(E^\mathrm{sta}_\Lambda)$ for each state $\Lambda$ of interest.
This is not practical for simulations of spectra, where a large number of eigenstates $N_\Lambda$ is required.
In this work, we instead group the $E^\mathrm{sta}_\Lambda$ into energy intervals of 0.3 eV  which allows us to reduce the number of times we need to evaluate the screening matrix $\tilde{W}(E^{\mathrm{sta}}_{\Lambda})$ from $N_{\Lambda}$ to the number of chosen frequency intervals.
For eigenstate $\Lambda$ with an eigenvalue $E^{\text{sta}}_{\Lambda}$ in a given interval, the dynamical screening function is approximated by the lower end of the interval.
In Sec.\ \ref{sec:res} of this work, the number of screening potential evaluations needed to compute dynamical corrections to the energies and spectra is reduced from 10$^4$ to about 30.

In addition, we note that in practice even for static screening the full diagonalization is usually avoided, using e.g.\ a time propagation approach \cite{PhysRevLett.88.016402,PhysRevB.67.085307}.
This would be feasible in the context of this work only by using the effective static approach discussed in the following.
In this approach, effective, static screening can be adopted to obtain an approximate solution of Eq.\ \eqref{eq:solution}. 
This bypasses the frequency-dependent eigenvalue problem entirely, but instead focuses on approximating Eq.\ \eqref{equ:w} by replacing the energy difference terms, $E_{c'\mathbf{k}'}-E_{v\mathbf{k}}-\hbar \omega$ and $E_{c\mathbf{k}}-E_{v'\mathbf{k}'}-\hbar \omega$ by an effective, constant exciton-binding energy, that is independent of the energies of the electronic states and $\hbar\omega$.
This reduces the dynamical screening problem to an effectively static problem since the two terms in the brackets of Eq.\ \eqref{equ:w} reduce to one single value, that can be chosen as the binding energy of the lowest excitonic state
\begin{equation}\label{eq:eb}
E_b=E_g-E^\mathrm{sta}_{\Lambda=0},
\end{equation}
where $E_g$ is the band gap without considering excitonic effects.
As a result, this approach replaces the dynamical screening function with an effective static screening function, that takes the exciton binding energy of the material explicitly into consideration and Eq.\ \eqref{equ:w} is simplified to\cite{gao2016dynamical}
\begin{widetext}
\begin{equation}
\label{equ:weff}
\begin{split}
\tilde{W}^{\text{eff}}_{cc',vv',\mathbf{kk}'}=\frac{1}{V} %\sum_{\mathbf{q}}
&\sum_{\mathbf{q, G, G'}}v\left(\sqrt{|\mathbf{q}+\mathbf{G}||\mathbf{q}+\mathbf{G'}|}\right)B_{cc',\mathbf{kk}'}(\mathbf{q}+\mathbf{G})B_{vv',\mathbf{kk}'}^{*}(\mathbf{q}+\mathbf{G'}) \\
&\times \left\{\delta_{\mathbf{GG'}}+\int_0^{\infty}\frac{d\hbar \omega '}{\pi}\text{Im}\varepsilon^{-1}(\mathbf{q}+\mathbf{G},\mathbf{q}+\mathbf{G'},\omega ')\frac{2}{\hbar\omega '+E_b}\right\}\delta_{\mathbf{q},\mathbf{k-k'}}.
\end{split}
\end{equation}
\end{widetext}

The resulting Eq.\ \eqref{equ:weff} contains no frequency dependence anymore since $\omega'$ can be integrated explicitly.
This approach is the cheapest among the three, as it is a modified version of the static approximation and it has been used to study effects of free-carrier screening\cite{gao2016dynamical} and the screening of lattice polarizability \cite{PhysRevLett.127.067401}.
In addition, among the three approaches we introduced, the effective static screening does not require the excitonic wavefunction, allowing us to take advantage of the time propagation approach\cite{PhysRevLett.88.016402,PhysRevB.67.085307} to avoid the diagonalization of the eigenvalue problem.
In Ref.\ \onlinecite{gao2016dynamical}, the authors argue that the process can be repeated several times to converge the solution. 
However, we note that it needs to be tested whether the converged values will match the true solution of the frequency-dependent BSE.

\subsection{\label{sec:dynscreen}The dynamical screening function}

In order to proceed with solving the dynamical eigenvalue problem, Eq.\ \eqref{eq:bsefreq}, one needs to compute the \emph{frequency-dependent} screened Coulomb interaction, Eq.\ \eqref{equ:w}.
The major challenge lies in the frequency integral with respect to $\omega'$.
While it can be evaluated numerically, e.g.\ within the random-phase approximation \cite{ren2012random,onida2002electronic}, this comes with high computational cost, since one $\omega'$ integral needs to be evaluated explicitly for each $\omega$ and each combination of $cv\mathbf{k}$ and $c'v'\mathbf{k'}$, see Eq.\ \eqref{equ:w}.

The integral can be carried out explicitly if an analytical model function is assumed for the $\omega'$ dependence of the inverse dielectric matrix.
In this work, we pursue that route and use the generalized plasmon-pole approximation (PPA) from Hybertsen and Louie \cite{hybertsen1986electron,rohlfing2000electron,botti2013strong} to carry out the frequency integral.
This model expresses the frequency dependent inverse dielectric matrix to be a pole function of the form
\begin{widetext}
\begin{equation}
\label{eq:imppa}
\text{Im}\,\varepsilon^{-1}(\mathbf{q+G},\mathbf{q+G'},\omega)=A(\mathbf{q+G},\mathbf{q+G'})\times \{ \delta[\omega-\tilde{\omega}(\mathbf{q+G},\mathbf{q+G'})]-\delta[\omega+\tilde{\omega}(\mathbf{q+G},\mathbf{q+G'})] \}
\end{equation}
\begin{equation}
\label{eq:reppa}
\text{Re}\,\varepsilon^{-1}(\mathbf{q+G},\mathbf{q+G'},\omega)=1+\frac{\Omega^2(\mathbf{q+G},\mathbf{q+G'})}{\omega^2-\tilde{\omega}^2(\mathbf{q+G},\mathbf{q+G'})}.
\end{equation}
\end{widetext}
The three parameters $A(\mathbf{q+G,q+G'})$,  $\tilde{\omega}(\mathbf{q+G,q+G'})$, and $\Omega(\mathbf{q+G,q+G'})$ %\as{You're switching from parentheses to indices in the notation?}
are given by three additional constraints, i.e.\ the Kramers-Kronig relation, the $f$-sum rule, and the \emph{static} inverse dielectric matrix $\varepsilon^{-1}(\mathbf{q+G},\mathbf{q+G'},\omega=0)$\cite{hybertsen1986electron}.
To describe the wave-vector dependence of the \emph{static} inverse dielectric matrix we adopt the model from Bechstedt \emph{et al.}\cite{bechstedt1992efficient}, which considers only diagonal terms $\mathbf{G}=\mathbf{G'}$. 
It interpolates between free-electron gas behavior at large $\mathbf{q}$ and Thomas-Fermi like behavior at small $\mathbf{q}$ \cite{bechstedt1992efficient,fuchs2008efficient,Roedl:2008}.
The approximation of neglecting local-field effects in the screening ($\mathbf{G=G'}$) is reasonable in typical semiconductors \cite{bechstedt1992efficient,schleife2009optical,schleife2011electronic}.
Whether it can impose problems for studying dynamical screening effects, e.g.\ when excitons become more localized, remains worthwhile exploring in the future.

With the above constraints the three parameters of the plasmon-pole model follow as (see the supplemental information section B for details of the derivation\cite{supplement})
\begin{eqnarray}
\label{equ:omegalf}
\Omega(\mathbf{q+G})&=&\omega_p,\\
\label{equ:tomegalf}
\tilde{\omega}(\mathbf{q+G})&=&\omega_p[1-\varepsilon^{-1}(\mathbf{q+G},\omega=0)]^{-1/2},\\
\label{equ:alf}
A(\mathbf{q+G})&=&-\frac{\pi}{2}\omega_p[1-\varepsilon^{-1}(\mathbf{q+G},\omega=0)]^{1/2}.
\end{eqnarray}
Carrying out the frequency integral in Eq.\ \eqref{equ:w} then provides the dynamical screening potential
\begin{widetext}
\begin{equation}
\label{eq:wbloch}
\begin{split}
\tilde{W}_{cc',vv',\mathbf{kk}'}(\omega)=&\frac{1}{V} \sum_{\mathbf{q, G}}v(|\mathbf{q}+\mathbf{G}|)B_{cc',\mathbf{kk}'}(\mathbf{q+G})B_{vv',\mathbf{kk}'}^{*}(\mathbf{q+G}) \\
&\times \Bigg\{1-\frac{\hbar\omega_p}{2}\left[1-\varepsilon^{-1}(\mathbf{q+G},\omega=0)\right]^{1/2}\\
&\times \Bigg[\frac{1}{\hbar\omega_p(1-\varepsilon^{-1}(\mathbf{q+G},\omega=0))^{-1/2}+E_{c\mathbf{k}}-E_{v'\mathbf{k}'}-\hbar \omega}\\&+\frac{1}{\hbar\omega_p(1-\varepsilon^{-1}(\mathbf{q+G},\omega=0))^{-1/2}+E_{c'\mathbf{k}'}-E_{v\mathbf{k}}-\hbar \omega}\Bigg]\Bigg\}\delta_{\mathbf{q},\mathbf{k-k'}}.
\end{split}
\end{equation}
\end{widetext}
In the effective static screening approach the energy differences in the denominator are replaced by the exciton binding energy, see Eq.\ \eqref{eq:eb}, resulting in
\begin{widetext}
\begin{equation}
\label{eq:wblochstatic}
\begin{split}
\tilde{W}^{\text{eff}}_{cc',vv',\mathbf{kk}'}=&\frac{1}{V} \sum_{\mathbf{q, G}}v(|\mathbf{q}+\mathbf{G}|)B_{cc',\mathbf{kk}'}(\mathbf{q+G})B_{vv',\mathbf{kk}'}^{*}(\mathbf{q+G}) \\
&\times \Bigg\{1-\frac{\hbar\omega_p}{2}\left[1-\varepsilon^{-1}(\mathbf{q+G},\omega=0)\right]^{1/2}\\
&\times \Bigg[\frac{2}{\hbar\omega_p(1-\varepsilon^{-1}(\mathbf{q+G},\omega=0))^{-1/2}+E_b}\Bigg]\Bigg\}\delta_{\mathbf{q},\mathbf{k-k'}}.
\end{split}
\end{equation}
\end{widetext}

The denominators in Eqs.\ \eqref{eq:wbloch} and \eqref{eq:wblochstatic} significantly determine the nature of screening through the interplay between the plasma frequency $\omega_p$ as a characteristic frequency, and exciton binding, either expressed as the two energy differences $E_{c'\mathbf{k}'}-E_{v\mathbf{k}}-\hbar \omega$ and $E_{c\mathbf{k}}-E_{v'\mathbf{k}'}-\hbar \omega$ in Eq.\ \eqref{eq:wbloch} or $E_b$ in Eq.\ \eqref{eq:wblochstatic}.
The static limit corresponds to negligible exciton binding compared to the plasma frequency and can be obtained from Eqs.\ \eqref{eq:wbloch} and \eqref{eq:wblochstatic} by dropping these energy differences or $E_b$, respectively.
In this case, all terms in the curly brackets reduce to $\varepsilon^{-1}(\mathbf{q+G},\omega=0)$.
In bulk semiconductors, plasma frequencies are usually several eV to several tens of eV and exciton-binding energies are several tens to a few hundreds of meV, i.e.\ at least one order of magnitude smaller, illustrating the validity of the static approximation.
In many low-dimensional or organic semiconductors, however, the exciton binding energies are relatively large and can be on the order of 1 eV\cite{das2018electronic,PhysRevB.53.15909,hummer2005oligoacene}, rendering the validity of the static approximation questionable.
Furthermore, this illustrates, e.g.\ for the lowest bound excitonic state, that including electronic screening dynamics effectively reduces screening compared to the static approximation, leading to stronger excitonic effects.
This is because the denominator of Eq.\ \eqref{eq:wbloch} is larger than when $E_b$ is dropped in the static case.
Hence, dynamical screening is effectively weaker and in the static approximation screening is always overestimated.
Physically, this can be interpreted as an initially incomplete screening in the dynamic case, compared to an instantly formed screening in the static approximation \cite{bechstedt2016many}.

\section{\label{sec:comp}Computational methods}

In this work we compare the three different approaches to describe screening dynamics, i.e.\ exact diagonalization, perturbative treatment, and the effective static screening approach, for optical spectra and exciton binding energies of crystalline naphthalene.
This material is an organic crystal for which large exciton binding energies of 1.0\,--\,1.5 eV were reported from experiment \cite{hummer2005oligoacene}.
We implemented the three different approaches using the plasmon-pole approximation (PPA), into the BSE implementation discussed in Refs.\ \onlinecite{Roedl:2008,fuchs2008efficient}, based on the Vienna \emph{Ab-Initio} Simulation Package \cite{Gajdos:2006,Kresse:1999,kresse96} (VASP).

For naphthalene, we first performed density functional theory \cite{PhysRev.140.A1133,PhysRev.136.B864} (DFT) simulations using the generalized-gradient approximation (GGA) by Perdew, Burke, and Ernzerhof (PBE) to describe exchange and correlation \cite{PhysRevLett.77.3865} and the projector-augmented wave (PAW) scheme \cite{blo94} to model the electron-ion interaction.
Kohn-Sham states were expanded into plane waves up to a cutoff energy of 400 eV.
We used lattice constants that were reported from experiment \cite{capelli2006molecular} and relaxed  atomic positions until all Hellmann-Feynman forces were smaller than 5 meV/\r{A}, using the DFT-D2 method of Grimme\cite{grimme2006semiempirical} to capture Van der Waals corrections.
For these relaxations, the Brillouin zone (BZ) was sampled using $3\times5\times3$ $\Gamma$-centered  %\as{Put MP or Gamma centered here.}
$\mathbf{k}$ points.
We verified that the total energy of the unit cell is converged to better than 1 meV/atom with these parameters. 

For the BSE simulations, we computed the DFT-PBE electronic structure for the relaxed atomic geometries described above and tested convergence with respect to BZ sampling and BSE cutoff energy, i.e.\ the energy up to which non-interacting electron-hole pairs are included in the BSE Hamiltonian.
We did these tests using static screening and all details can be found in the supplemental information\cite{supplement} (see Figs. S4, S5, and S6).
We find that, contrary to materials with dispersive valence and conduction bands such as MgO and ZnO \cite{fuchs2008efficient,zhang2018nonequilibrium}, the valence and conduction band edges of naphthalene are flat (see Fig. S1), and the exciton binding energy of naphthalene converges quickly with BZ sampling (see Fig. S5).
We obtain the value of the exciton binding energy by extrapolating to infinitely dense sampling as discussed in Ref.\ \onlinecite{fuchs2008efficient}. 
Balancing computational cost and accuracy of the BSE calculations of optical spectra, we adopted a $5\times7\times5$ $\mathbf{k}$-point grid centered at the $A$ point of the BZ, to capture the lowest-energy transitions near that point.
We use a BSE cutoff energy of 14 eV to compute spectra with static screening and compare these to literature results in Fig.\ \ref{fig:comp_gga}.
Due to the larger computational cost of dynamical screening, we reduce the BSE cutoff energy to 9 eV for investigating dynamical effects, and focus on the spectra between the onset at 3 eV and up to 5.5 eV. 
Based on the convergence tests above, we anticipate that the choice of the energy cutoff and $\mathbf{k}$-points sampling results in deviation around 0.2 eV compared to the converged values, however, we show in the SI Sec. G that these error induces constant shifts to the predicted exciton binding energy, and we do not expect them to affect our analysis of dynamical screening effects.
In all spectra calculations, a Lorentzian life-time broadening of 0.1 eV is used.
In the static model dielectric function, a high-frequency dielectric constant of 2.35 is used, and is chosen based on the experimental value \cite{wohlfarth2008static,suthan2010growth}.

\section{\label{sec:res}Results and discussion}

\begin{figure}
\includegraphics[width=0.45\textwidth]{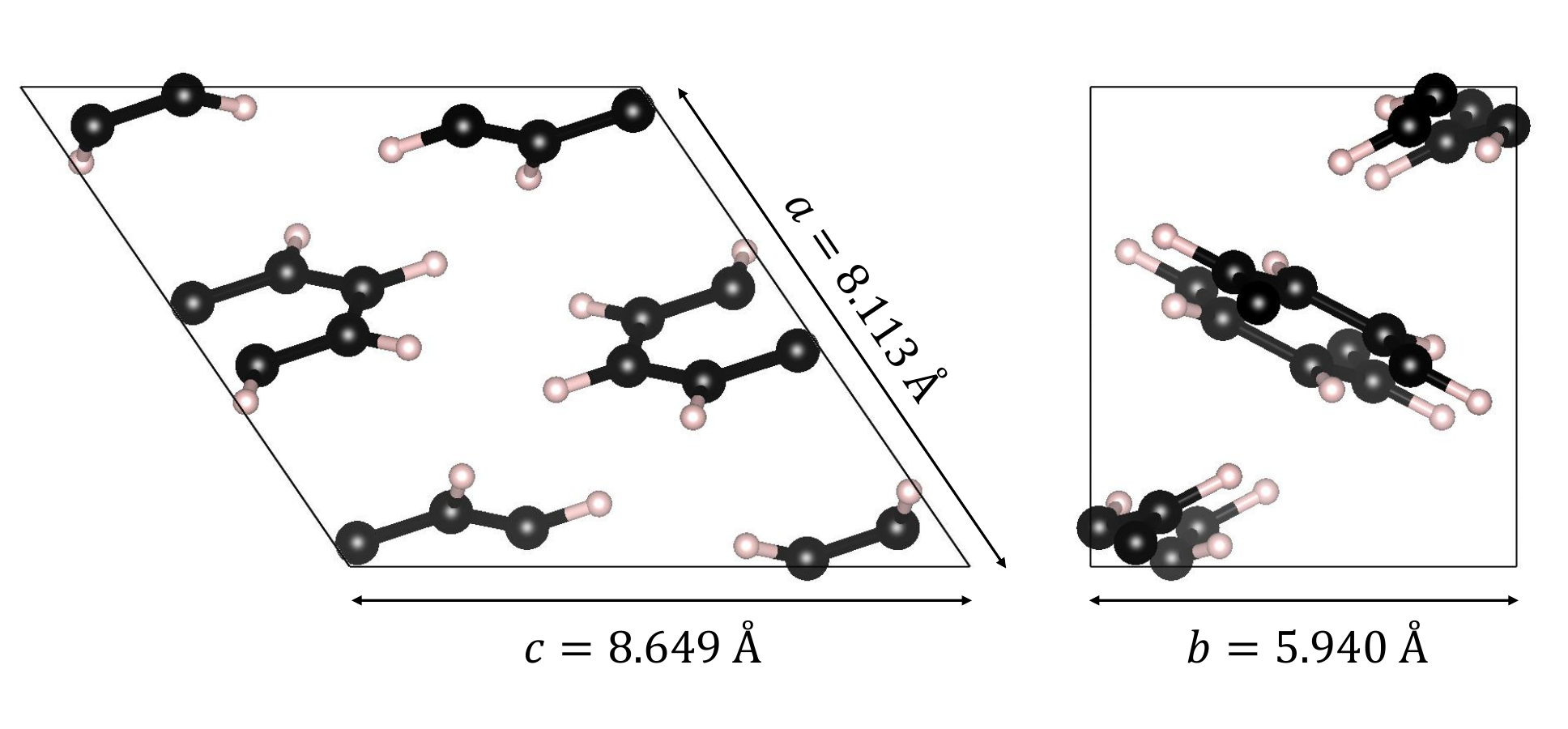}
\caption{\label{fig:crystal}(Color online.)
Monoclinic naphthalene (C$_{10}$H$_8$) viewed from the crystalline $b$ direction (left) and the crystalline $c$ direction (right). 
Black spheres represent carbon atoms and white spheres represent hydrogen atoms. 
The unit cell consists of two conjugated orientated naphthyl rings. 
The crystal structure is obtained from Ref.\ \onlinecite{capelli2006molecular} and we subsequently fully relaxed all atomic positions.
}
\end{figure}

We compute the optical spectra and exciton binding energies of the organic crystal naphthalene using static screening in the BSE and the three different approaches to dynamical electronic screening discussed above.
The unit cell of naphthalene is shown in Fig.\ \ref{fig:crystal} and consists of two units of double carbon rings with conjugated orientation. 
This system is an ideal test bed to systematically study dynamical effects in the description of electronic screening, since it exhibits exciton binding energies on the order of 1 eV \cite{hummer2005oligoacene,hummer2005electronic,pope1999electronic}, which is an order of magnitude larger than exciton binding energies of typical bulk inorganic semiconductors of several 10 meV \cite{yu1996fundamentals,roessler1967electronic,whited1973exciton}.
Using the independent-particle approach in VASP and the integral from the $f$-sum rule without considering the electron-hole interaction, we compute a plasma frequency of 17.9 eV, which is similar to bulk inorganic materials.
The closer the exciton binding energy is to the plasma frequency, the more important are dynamical effects for electronic screening, and this is what we expect for naphthalene in this work.

\subsection{\label{sec:gga_sta}Independent quasi-particle approximation and static BSE}

\begin{figure}
\includegraphics[width=0.45\textwidth]{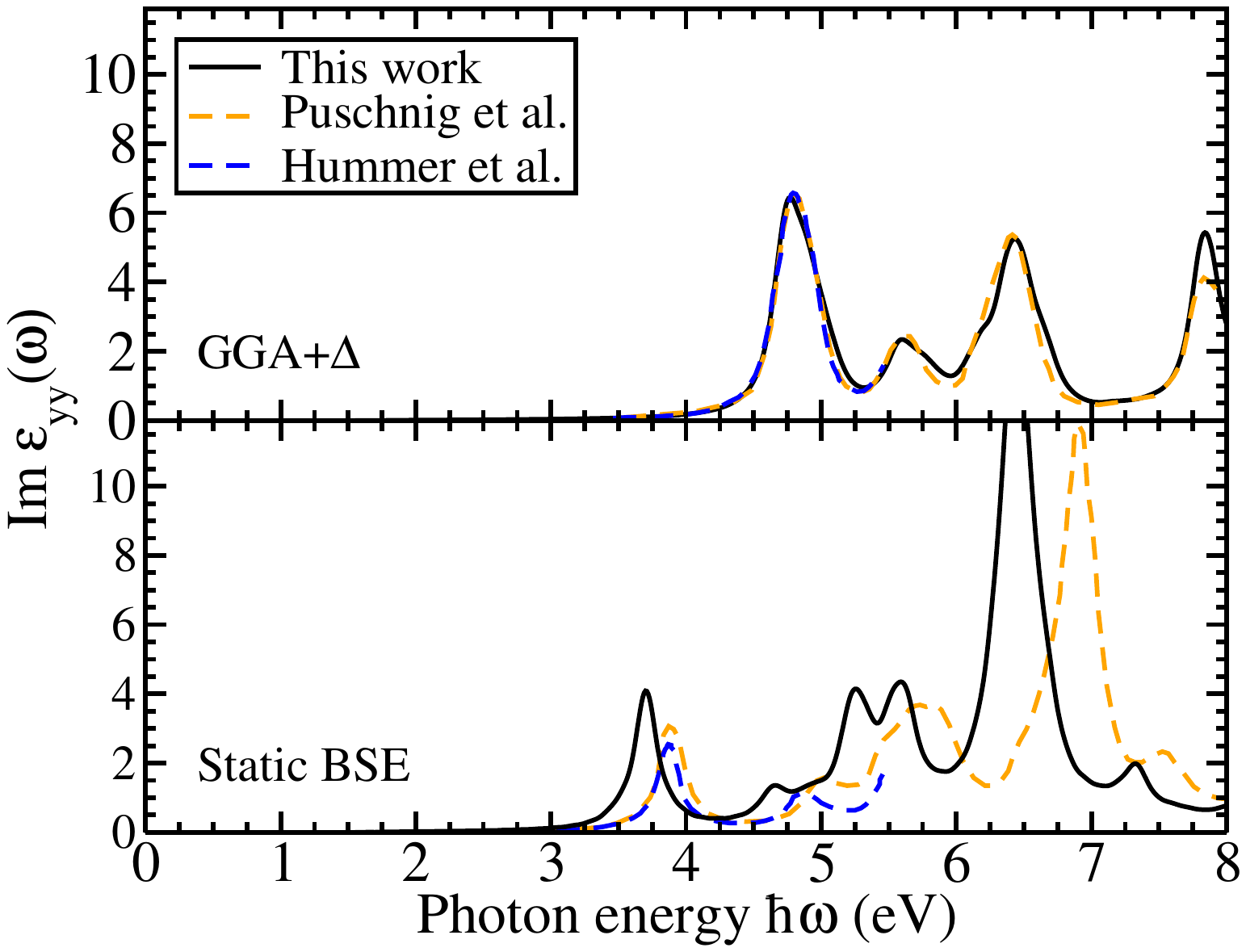}
\caption{\label{fig:comp_gga}(Color online.)
Imaginary part of the $\varepsilon_{yy} \parallel b$ component of the dielectric tensor, calculated without electron-hole interaction (upper panel) and from solving the BSE with statically screened electron-hole interaction (lower panel). 
We used the high-frequency dielectric constant from experiment\cite{wohlfarth2008static,suthan2010growth}, $\varepsilon_{\infty}$=2.35, and a scissor shift of $\Delta$=1.55 eV.
The shape of our spectra agrees very well with data by Puschnig \emph{et al.}\cite{puschnig2013excited} and Hummer \emph{et al.}\cite{hummer2005oligoacene}
}
\end{figure}

We first compute the optical spectrum of naphthalene using the independent-quasiparticle approximation within the GGA+$\Delta$ approach as well as the static BSE, see Fig.\ \ref{fig:comp_gga}.
In this work, we focus on optical spectra and exciton binding energy for the $y$-polarization, i.e.\ the $\varepsilon_{yy}$ component of the dielectric tensor parallel to the crystalline $b$ direction, since for this direction the lowest-energy excitonic eigenstates were reported \cite{hummer2005electronic,hummer2005oligoacene}. 
Our calculated value for the GGA band gap, $E_\mathrm{g}^\mathrm{GGA}$=3.12 eV, agrees well with an earlier result of 3.10 eV \cite{hummer2005electronic}. 
We use a scissor shift of $\Delta$=1.55 eV so that the first bright peak is at the same position at 4.8 eV as reported from quasiparticle calculations \cite{hummer2005oligoacene}.
Details of the band structure can be found in Fig. S1 of the supplemental information\cite{supplement}.
The upper panel of Fig.\ \ref{fig:comp_gga} shows good agreement between our GGA+$\Delta$ result and a spectrum from the literature \cite{hummer2005oligoacene,puschnig2013excited}.
We notice that some differences are observed at the peak around 8 eV, as the height of the peak from our calculation is slightly larger.
We also verified that our results agree very well with a GGA+$\Delta$ spectrum using a broadening of 0.05 eV\cite{hummer2005electronic}, if we adopt the same broadening (see supplemental information\cite{supplement} Sec. D).

When including the electron-hole interaction by solving a BSE with static screening, we find strong excitonic effects in naphthalene and indeed report an exciton-binding energy of 1.06 eV.
The difference between the first main peak with (BSE) and without (GGA+$\Delta$) excitonic effects, see upper and lower panel of Fig.\ \ref{fig:comp_gga}, is used to obtain the value of the exciton binding energy, similar as in Ref.\ \onlinecite{hummer2005oligoacene}.
Comparing our static BSE spectrum with earlier results in the literature in the lower panel of Fig.\ \ref{fig:comp_gga} shows reasonable agreement also of the overall spectral shape  \cite{hummer2005oligoacene,puschnig2013excited}. 
We note that Puschnig \emph{et al.}\ used a broadening of 0.2 eV\cite{puschnig2013excited}, possibly explaining some of the deviations with respect to our data.
In addition, our predicted exciton binding energy is slightly larger than 0.9 eV reported in Ref.\ \onlinecite{hummer2005oligoacene}, and correspondingly, the onset of our spectrum in Fig.\ \ref{fig:comp_gga} appears at slightly lower energy.
We attribute this to the different approaches of describing the wave-vector dependence of the screening the static BSE.
While our work uses the Bechstedt model and the experimental high-frequency dielectric constant of 2.35\cite{wohlfarth2008static,suthan2010growth}, Ref.\ \onlinecite{puschnig2013excited} and Ref.\ \onlinecite{hummer2005oligoacene} use the RPA based on the Kohn-Sham eigenvalues\cite{PhysRevB.66.165105}, and the corresponding dielectric constant computed within the independent particle approximation is reported in Ref. \ \onlinecite{puschnig2013excited} as 3.8.

\subsection{\label{sec:dyn}Dynamical electronic screening}

\begin{figure}
\includegraphics[width=\columnwidth]{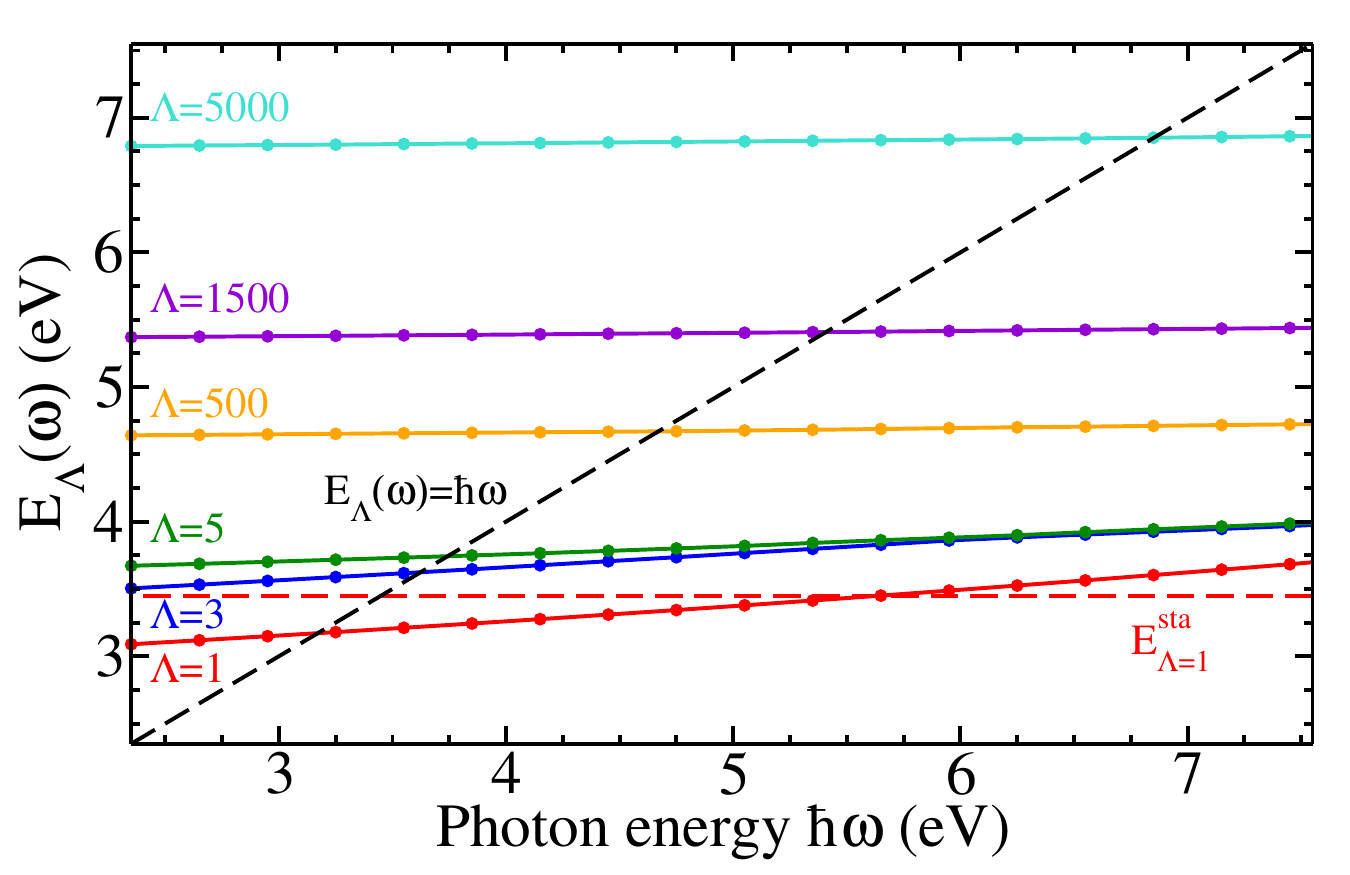}
\caption{\label{fig:illu}(Color online.)
Frequency dependent exciton eigenvalues $E_\Lambda(\omega)$ for excitonic states $\Lambda$, obtained using exact diagonalization of the Hamiltonian in Eq.\ \eqref{eq:bsefreq}.
We show various randomly selected states to cover a range of exciton energies.
We find the solution of $E_\Lambda(\omega)=\hbar\omega$ (black dashed line) via a nearest neighbor approximation (see text).
The solution using the standard static approximation for the lowest excitonic state $(\Lambda=1)$ is marked with the horizontal red-dashed line. 
}
\end{figure}

We now compare the spectra we computed from solutions of the BSE that account for electronic screening dynamics via the three different approaches discussed in Sec.\ \ref{sec:theory}.
First, we compute excitonic eigenvalues $E_\Lambda$ via direct diagonalization by sampling a frequency grid to solve Eq.\ \eqref{eq:solution}.
Figure \ref{fig:illu} shows this sampling of the frequency range of interest with a spacing of 0.3 eV and our computed solutions for the excitonic eigenvalues $E_\Lambda$.
This figure shows that there is no complicated dependence on frequency, illustrating that a simple interpolation scheme is appropriate and our frequency spacing of 0.3 eV is sufficient.
We further verified this by calculating spectra using different samplings, and the results can be found in Fig. S3 of the supplemental information\cite{supplement}.
Here we use a nearest neighbor approximation, i.e.\ for each state, the solution that is the closest to the $E_\Lambda(\omega)=\hbar\omega$ line is adopted as the solution of the dynamical problem for that state. 
This allows us to also compute optical spectra, which requires excitonic wave functions that would be more challenging to obtain in an interpolation scheme.
We estimate from Fig.\ \ref{fig:illu} that the nearest neighbour approximation does not cause an error in the solution of more than 0.01 eV.
Nevertheless Fig.\ \ref{fig:illu}  illustrates, e.g.\ for the $\Lambda=1$ state, that an exciton binding energy of 1.49 eV compared to the plasma frequency of 17.9 eV is affected by the frequency dependence due to electronic screening dynamics.

\begin{figure}
\includegraphics[width=0.48\textwidth]{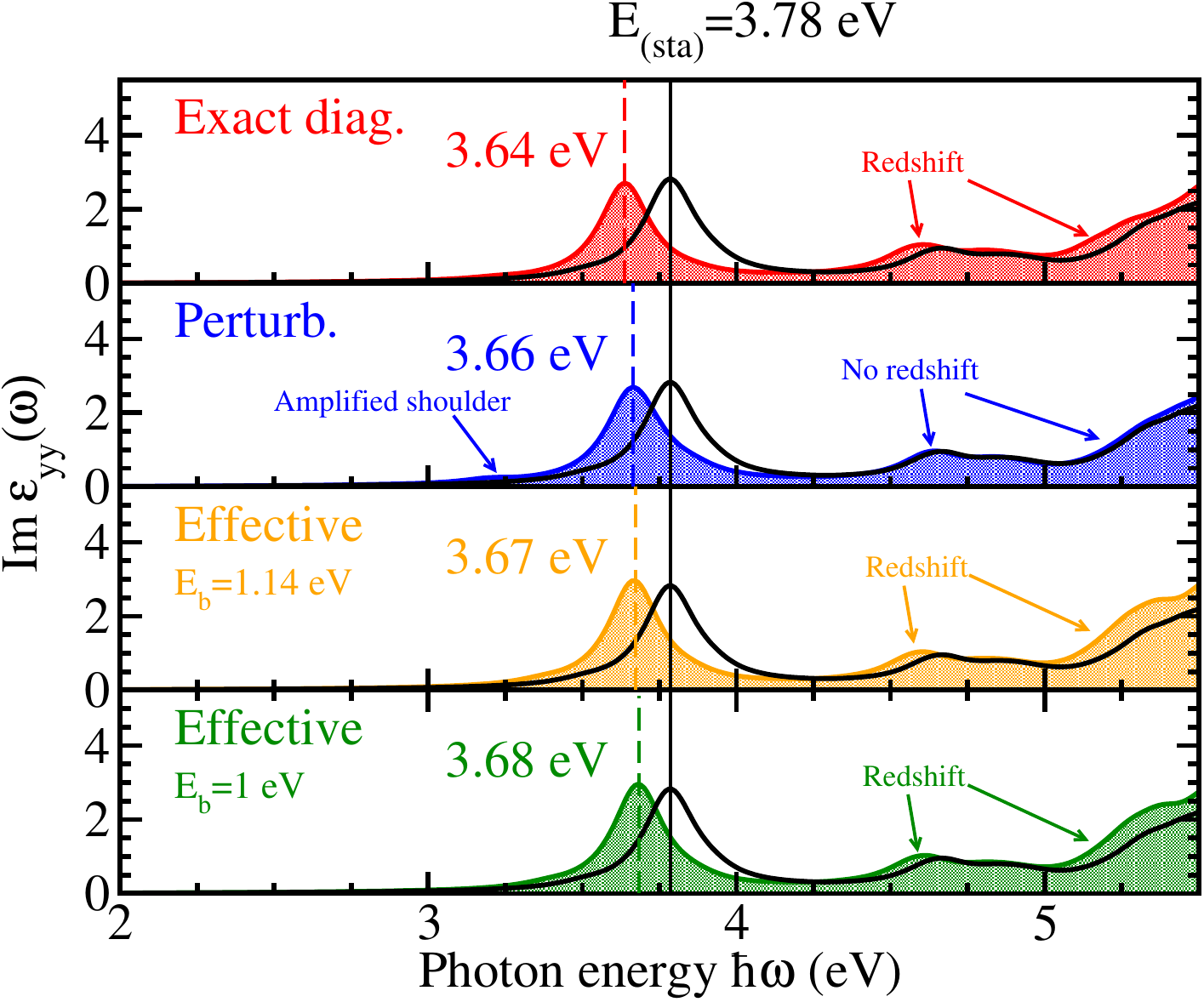}
\caption{\label{fig:spec}(Color online.)
Comparison of $\varepsilon_\text{yy}$ from the static BSE (black solid lines) with the different approaches to include dynamical screening.
Red shows the exact diagonalization and blue results from the perturbative approach.
Effective static screening using $E_b=1.14$ eV (orange) and $E_b=1$ eV (green) is also shown.
All curves use a scissor shift of 1.55 eV.
The positions of the first major peak are highlighted in the figure with vertical dashed lines.
We also note that an excitonic state is visible when adopting the perturbative approach, that is dark in all other cases.
}
\end{figure}

In the upper two panels of Fig.\ \ref{fig:spec} we compare the optical spectra from the exact diagonalization of the dynamical problem and the perturbative approach to our static BSE result. 
This comparison shows that the perturbative treatment works well for the lowest-energy major peak predicted by the static approximation at 3.78 eV, as it predicts a 0.12 eV redshift from the peak with static screening, compared to a redshift of 0.14 eV when the exact diagonalization approach is used.
Figure \ref{fig:spec} also shows a redshift of all spectra that include screening dynamics relative to the static case, which confirms the expected effective reduction of screening when dynamics is included, as discussed in Sec.\ \ref{sec:dynscreen}.

Further comparison shows that the perturbative treatment results in a magnified excitonic shoulder at a photon energy of 3.25 eV, see Fig.\ \ref{fig:spec}, which neither of the other techniques shows. We find a corresponding excitonic eigenvalue using all three approximations to dynamical screening, however, its oscillator strength is much smaller in the case of exact diagonalization and for the effective static treatment of screening. Inclusion of screening dynamics reduces the oscillator strength of the underlying excitonic states. The perturbative approach struggles with this, as it uses unchanged excitonic eigenfunctions of the static simulation. In addition, at higher energies beyond about 4.2 eV, we see that the perturbative treatment strongly resembles the static result, while all other approaches yield an overall redshift and an enhancement of the shoulder near 5 eV (see Fig. \ref{fig:spec}). 

Finally, we investigate the effective static screening approximation, see Eq.\ \eqref{eq:wblochstatic}.
In Fig.\ \ref{fig:spec} we show results for effective static screening using the exciton-binding energy calculated from the static screening approximation (1.0 eV) and that from exact diagonalization (1.14 eV), leading to dynamical screening corrections of 0.10 eV and 0.11 eV, respectively.
The lower two panels of Fig.\ \ref{fig:spec} show that both cases underestimate the correction due to dynamical screening compared to the exact diagonalization.
The position of the first peak is about 0.03\,--\,0.04 eV higher in energy, i.e.\ closer to the static approach, corresponding to a 28.5\,\% and 21.4\,\% change of the dynamical correction to the exciton binding energy when compared to exact diagonalization.
Overall, however, effective static screening describes spectra better than the perturbative approach especially at higher energies as the perturbative treatment fails to capture the redshift due to dynamical effects qualitatively. 
As a result, we recommend the effective static approximation over the perturbative approach to include dynamical screening when optical spectra over a range of photon energies are considered. 

We provide an analysis to understand the difference we observed above between the different approaches: Including screening dynamics either via exact diagonalization or via the effective static screening approach changes how individual single-particle Kohn-Sham states are combined into excitonic wave functions. This is because the matrix elements of the BSE Hamiltonian change and the eigenstate of that Hamiltonian determines the exciton wave function. Such a change of single-particle KS contributions can affect how strongly or weakly a certain feature appears in the spectrum. In the case of the perturbative approach, see Eq. (11), the exciton wave functions themselves are kept constant and mixing of individual KS states is not changed, relative to static screening BSE simulations.

We note that Hummer \emph{et al.}\ \cite{hummer2005oligoacene} report a minor underestimation of the exciton-binding energy compared to experiment.
Their theoretical value is 0.9 eV, while experimental results are reported as 1.0\,--\,1.5 eV  \cite{hummer2005oligoacene}.
We find that the reduction of the screening due to dynamical effects provides an additional redshift to the spectra, putting the predicted exciton-binding energy closer to the range of the experimental values. This is additional evidence that electronic dynamical screening effects need to be taken into consideration for accurate modeling of these systems where large exciton-binding energies are observed. 
We note that while lowest singlet excitation energies reported from experiment are reasonably consistent (3.9 - 4.0 eV\cite{braun1970intrinsic,bergman1974two,swiderek1990electron,hanson1969information}), a variation of the reported electronic gap (5.0 - 5.4 eV\cite{SATO1987269,belkind1974,braun1970intrinsic,riga1977comparative}) is seen, due to the models used to extract the gap from photoemission experiments. This causes a variation of the estimated binding energy from experiments on the order of 0.5 eV. Even if we exclude the electronic band gap of 5.0 eV reported in Ref.\ \onlinecite{braun1970intrinsic} from observations of the change in carrier generation, the remaining references still show a variation of 5.1-5.4 eV with an error bar on the order of 0.2 eV due to uncertainty in empirical parameters\cite{riga1977comparative}, resulting in an estimated exciton-binding energy of 1.2-1.5 eV.

\subsection{\label{sec:omegap}Influence of the plasmon-pole model}

\begin{figure}
\includegraphics[width=0.98\columnwidth]{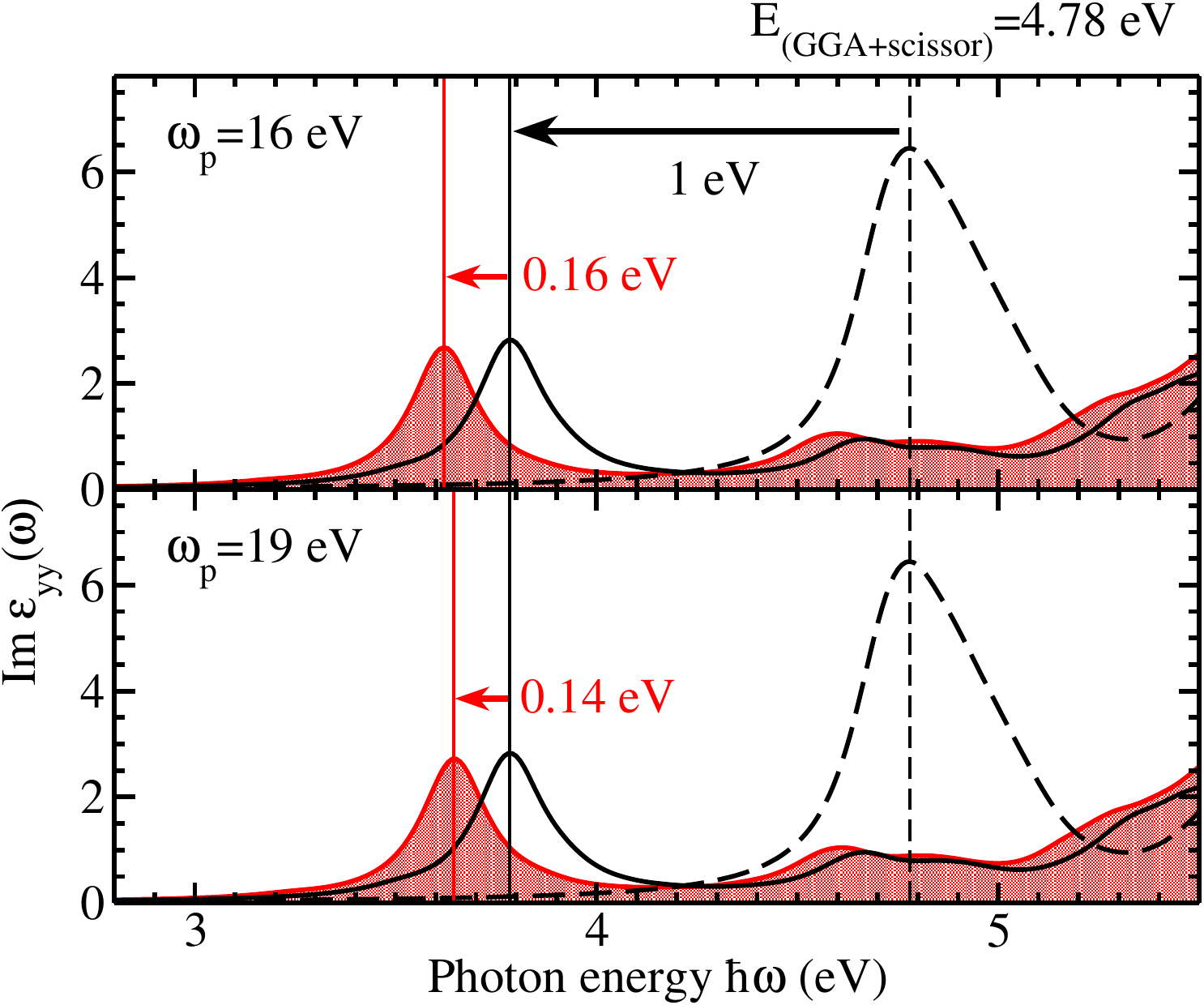}
\caption{\label{fig:plsm}(Color online.)
Imaginary part of the $\varepsilon_{yy} \parallel b$ component of the dielectric tensor from 
static BSE (black solid line) and independent quasiparticle approximation (GGA+scissor, black dashed line) are compared to exact diagonalization results computed using different values of the plasmon frequency (red shaded).
We show that increasing $\omega_p$ from 16 eV to 19 eV changes the prediction of the correction on the exciton binding energy due to dynamical screening effects from 0.16 eV to 0.14 eV. 
}
\end{figure}

As discussed in Sec.\ \ref{sec:dynscreen}, in this work a plasmon-pole model is adopted to derive Eq.\ \eqref{eq:wbloch}, which requires the plasma frequency $\omega_p$ as an input to calculate the screened Coulomb potential.
We compute $\omega_p$=17.9 eV within independent-particle approximation, i.e.\ without considering the electron-hole interaction, through integrating the imaginary part of the dielectric function  using the VASP code and averaging over the Cartesian coordinates. 
It has been found previously that the enforcement of the $f$-sum rule of the HL PPM model can overestimate the energy of the pole of the dielectric function, e.g.\ by about 10\,\% for elemental carbon \cite{larson2013role}.

Hence, in the following we examine the influence of the plasmon-pole frequency $\omega_p$ on our results and varied $\omega_p$ in a $\sim$10\,\% range from the calculated value, between 16 and 19 eV, to examine the influence on resulting spectra and exciton-binding energy.
From the spectra shown in Fig.\ \ref{fig:plsm} we see that with increasing plasma frequency from 16 eV to 19 eV, the first major peak is slightly blue shifted because the predicted red shift due to dynamical screening corrections is increased from 0.14 eV to 0.16 eV, i.e.\ by about 15\,\%.
This reduction of the dynamical correction is expected, since increasingly large $\omega_p$ corresponds more closely to the static screening case, as discussed in Sec.\ \ref{sec:dynscreen}, reducing dynamical corrections and increasing the strength dielectric screening.
However, the influence is not substantial around the plasma frequency of interest in this work.
Overall, we found that the potential overestimation of the pole energy due to the choice of the plasmon-pole model, does not qualitatively affect the significance of dynamical effects. 
We also note that the dependence on the plasmon-pole energy is similar to that of the effective static method and a similar trend is expected in this case.

Finally, we note that for naphthalene, the exciton binding energy is only 5\% of the plasma frequency. 
However, the ratio between exciton binding energy and characteristic frequency that determines screening dynamics is much larger in other important applications of the BSE technique.
For example, in the case of phonon screening in polar materials, the characteristic frequency scale to compare to is that of phonon frequencies, instead of the plasma frequency \cite{PhysRevLett.127.067401}.
In this case, the exciton-binding energy is on the same order compared to the characteristic frequency, and screening dynamics is expected to be important for an accurate description of optical properties.

\section{\label{sec:conclusion}Conclusion}

We examined the effects of electronic dynamical screening of the electron-hole interaction when solving the Bethe-Sapleter equation of the optical polarization function for a naphthalene organic crystal.
By adopting the Shindo approximation and a plasmon-pole model, the BSE can be written as a frequency-dependent eigenvalue problem.
We compared three different ways of addressing this dynamical problem, i.e., exact diagonaliztion of the frequency-dependent BSE Hamiltonian, perturbative treatment of the static eigenstates, and an effective static approximation.
The exact-diagonalization approach requires solving the BSE Hamiltonian at numerous frequencies to obtain the eigenenergies and eigenstates. 
The perturbative approach requires diagonalization of the BSE Hamiltonian in the static approximation in order to obtain the excitonic wavefunctions. 
Meanwhile, the effective screening approach bears the same cost of the standard approach of solving the BSE within the static approximation. 

We show that for naphthalene, all three methods induce a $\sim$15\,\% correction of the exciton binding energy, predicted to be around 1 eV by the standard static approximation.
While the exact diagonalization constitutes a reference case in this work, it comes at high computational cost that renders this approach unfeasible for computing spectra.
The perturbative treatment is a decent alternative that does not require full solution of multiple BSE Hamiltonians while providing good qualitative estimates of binding energies of the lowest excitonic states. 
Finally, we show that for spectra, the effective static screening approach is well suited and numerically efficient, possible allowing application to complex materials.
The results for naphthalene are in good agreement with experiments.
We also note that these insights will have implications when lattice screening is considered, since then the characteristic frequency of phonons is close to the exciton-binding energy, likely exacerbating the importance of screening dynamics.
In this case, the validity of the perturbative treatment and effective screening approach is expected to be more questionable, and requires further investigation. To this end, our findings suggest that dynamical screening effects in these contexts remain a subject of future study.

\begin{acknowledgements}
We thank Dr. Felipe H. da Jornada, Dr. Steven G. Louie, and Dr. Emmanouil Kioupakis for fruitful discussions.
This material is based upon work supported by the National Science Foundation under Grant No.\ DMR-1555153.
This research is part of the Blue Waters sustained-petascale computing project, which is supported by the National Science Foundation (awards OCI-0725070 and ACI-1238993) and the state of Illinois.
Blue Waters is a joint effort of the University of Illinois at Urbana-Champaign and its National Center for Supercomputing Applications.
This work made use of the Illinois Campus Cluster, a computing resource that is operated by the Illinois Campus Cluster Program (ICCP) in conjunction with the National Center for Supercomputing Applications (NCSA) and which is supported by funds from the University of Illinois at Urbana-Champaign.
\end{acknowledgements}

\bibliography{ref}
\clearpage

\renewcommand\theequation{S\arabic{equation}}
\renewcommand\thefigure{S\arabic{figure}}

\setcounter{equation}{0} 
\setcounter{figure}{0} 
\onecolumngrid

\section*{Supplemental Information}

\subsection{\label{sup:shindo}Shindo's approximation}
In order to approximately include dynamical screening, Shindo's approximation is adopted, which expresses the two-frequency dependent polarization function in terms of the Green's function of non-interaction electrons and holes, as well as the one-frequency dependent polarization function \cite{bechstedt2016many,shindo1970effective}.
The approximation assumes that the two-frequency dependent polarization function takes the approximate form \cite{bechstedt2016many,shindo1970effective}

\begin{equation}
\label{eqapp:shindo}
P^M(\lambda_1\lambda'_1,\lambda_2\lambda'_2,z_nz_m)\approx \frac{G_{\lambda_1}(z_n)-G_{\lambda'_1}(z_n-z_m)}{-\frac{1}{i\hbar\beta}\sum_{n'}G_{\lambda_1}(z_{n'})-G_{\lambda'_1}(z_{n'}-z_m)}P^M(\lambda_1\lambda'_1,\lambda_2\lambda'_2,z_m).
\end{equation}

It can be shown that the denominator in Eq.\ \eqref{eqapp:shindo} 
can be simplified, in thermal equilibrium, to be frequency independent\cite{bechstedt2016many}, leading to a normalization factor $\tilde{N}_{\lambda_{1}\lambda_{1}^{\prime}}\left( z_m \right)$
\begin{equation}\label{eqapp:occup}
    \tilde{N}_{\lambda_{1} \lambda_{1}^{\prime}}\left( z_m \right)=
-\frac{1}{\beta} \sum_{n}\left[G_{\lambda_{1}}\left(z_{n}\right)-G_{\lambda_{1}^{\prime}}\left(z_{n}- z_m \right)\right]=f(\lambda_1')-f(\lambda_1).
\end{equation}

Shindo's approximation de-couples the frequency dependencies of the two-frequency dependent polarization function in Eq.\ \eqref{eq:fullbse} of the main text, with all the $z_n$ dependency described by the Green's functions for single particle states, as can be seen from Eq.\ \eqref{eqapp:shindo}.
Plugging Eq.\ \eqref{eqapp:shindo} and \eqref{eqapp:occup} into Eq.\ \eqref{eq:fullbse}, an expression of the single frequency dependent polarization function can be obtained \cite{bechstedt2016many}
\begin{equation}
\label{eqapp:polarizationdyn}
\begin{split}       P^M(\lambda_1\lambda'_1,\lambda_2\lambda'_2,z_m)=&\frac{-\tilde{N}_{\lambda_{1}\lambda_{1}^{\prime}}}{E_{\lambda_{1}}-E_{\lambda_{1}^{\prime}}-\hbar z_m}\times\{\delta_{\lambda_1\lambda'_2}\delta_{\lambda'_1\lambda_2}\\&+\sum_{\lambda_{3} \lambda_{4}}\left[\tilde{W}_{\lambda_{1}^{\prime} \lambda_{4}}^{\lambda_{1} \lambda_{3}}\left( z_m \right)+2 \overline{\mathrm{v}}_{\lambda_{3} \lambda_{4}^{\prime}}^{\prime}\right]  P^M(\lambda_3\lambda_4,\lambda_2\lambda'_2,z_m)\},
\end{split}
\end{equation}
Equation \eqref{eqapp:polarizationdyn} takes the same form as the usually used static limit, Eq.\ \eqref{eq:bse_static}, except for an effective dynamically screened Coulomb potential $\tilde{W}(z_m)$, 
\begin{equation}
\label{eqapp:dynw}
    \tilde{W}_{\lambda_{1}^{\prime} \lambda_{4}}^{\lambda_{1} \lambda_{3}}\left( z_m \right)=\frac{1}{\beta^{2}} \sum_{n, n^{\prime}} \frac{G_{\lambda_{1}}\left(z_{n}\right)-G_{\lambda_{1}^{\prime}}\left(z_{n}- z_m \right)}{\tilde{N}_{\lambda_{1} \lambda_{1}^{\prime}}\left( z_m \right)} W_{\lambda_{1}^{\prime} \lambda_{4}}^{\lambda_{1} \lambda_{3}}\left(z_{n}-z_{n^{\prime}}\right) \frac{G_{\lambda_{3}}\left(z_{n^{\prime}}\right)-G_{\lambda_{4}}\left(z_{n^{\prime}}- z_m \right)}{\tilde{N}_{\lambda_{3} \lambda_{4}}\left( z_m \right)}.
\end{equation}

We refer to Ref.\ \onlinecite{bechstedt2016many} for the procedure of carrying out the $n$ and $n'$ sums in Eq.\ \eqref{eqapp:dynw}, which results in the frequency integral of $\omega'$, see Eq.\ \eqref{equ:w}. 

\clearpage

\subsection{\label{app:ppm}Details of the dynamical screening function}

Next we show the derivation of the dynamical screening function, Eq.\ \eqref{eq:wbloch} in the main text. 
We start from the general form of the dynamical screening function derived from adopting Shindo's approximation, Eq.\ \eqref{eqapp:dynw}.
The frequency integral in the equation needs to be evaluated in order to obtain the dynamical screening function. 
As mentioned in the main text, we adopted the plasmon-pole model from Hybertsen and Louie \cite{hybertsen1986electron}. 
This model approximates the inverse dielectric function of the system with a single plasmon pole.
In the limit of considering only diagonal elements in $\mathbf{G}$, the inverse dielectric function is given by
\begin{eqnarray}
\label{eqapp:ppmimag}
\text{Im}\,\varepsilon^{-1}(\mathbf{q+G},\omega)&=&A(\mathbf{q+G})\{\delta[\omega-\tilde{\omega}(\mathbf{q+G})-\delta[\omega+\tilde{\omega}(\mathbf{q+G})]\}, \\
\label{eqapp:ppmreal}
\text{Re}\,\varepsilon^{-1}(\mathbf{q+G},\omega)&=&1+\frac{\Omega^2(\mathbf{q+G})}{\omega^2-\tilde{\omega}^2(\mathbf{q+G})}.
\end{eqnarray}
There are three parameters that need to be obtained, i.e.\ the amplitude $A(\mathbf{q+G})$, the pole frequency $\tilde{\omega}(\mathbf{q+G})$, and a term with units of frequency $\Omega(\mathbf{q+G})$. 
This model obtains these three quantities via three constraints. 
The first one is the Kramers-Kronig relation at $\omega=0$
\begin{equation}
\label{equapp:kk}
\text{Re}\,\varepsilon^{-1}(\mathbf{q+G},\omega=0)=1+\frac{2}{\pi}\int_0^\infty d\omega'\frac{1}{\omega'}\text{Im}\varepsilon^{-1}(\mathbf{q+G},\omega')
\end{equation}
which relates the real part of the dielectric function to the imaginary part.
Plugging Eq.\ \eqref{eqapp:ppmimag} and \eqref{eqapp:ppmreal} into Eq.\ \eqref{equapp:kk}, the following relationship is obtained
\begin{equation}
    1-\frac{\Omega^2(\mathbf{q+G})}{\tilde{\omega}^2(\mathbf{q+G})}=1+\frac{2}{\pi}\frac{A(\mathbf{q+G})}{\tilde{\omega}(\mathbf{q+G})},
\end{equation}
which simplifies to 
\begin{equation}\label{eqapp:kk_plugged}
A(\mathbf{q+G})=-\frac{\pi}{2}\frac{\Omega^2(\mathbf{q+G})}{\tilde{\omega}(\mathbf{q+G})}.
\end{equation}
The second constraint is given by the generalized $f$-sum rule \cite{hybertsen1986electron}
\begin{equation}
\label{equapp:omega}
\int_0^\infty d\omega \, \omega \text{Im}\varepsilon_{\mathbf{GG'}}^{-1}(\mathbf{q},\omega)=-\frac{\pi}{2}\omega_p^2\frac{(\mathbf{q}+\mathbf{G})\cdot(\mathbf{q}+\mathbf{G'})}{|\mathbf{q}+\mathbf{G}|^2}\frac{\rho(\mathbf{G}-\mathbf{G'})}{\rho(\mathbf{0})},
\end{equation}
which relates the imaginary part of the inverse dielectric matrix to the plasma frequency and the charge density in the crystal.
Considering only $\mathbf{G}=\mathbf{G'}$, the $f$-sum rule is simplified to
\begin{equation}
\label{equapp:omegag}
\int_0^\infty d\omega \, \omega \text{Im}\varepsilon^{-1}(\mathbf{q+G},\omega)=-\frac{\pi}{2}\omega_p^2.
\end{equation}
Plugging Eq.\ \eqref{eqapp:ppmimag} into \eqref{equapp:omegag} and carrying out the integral analytically, the following relationship is obtained
\begin{equation}\label{eqapp:fsum}
    A(\mathbf{q+G})\tilde{\omega}(\mathbf{q+G})=-\frac{\pi}{2}\omega_p^2.
\end{equation}
Finally, the third constraint is that Eq.\ \eqref{eqapp:ppmreal} at $\omega=0$ should correspond to the static inverse dielectric function, i.e.,
\begin{equation}\label{eqapp:static}
\varepsilon^{-1}(\mathbf{q+G},\omega=0)=1-\frac{\Omega^2(\mathbf{q+G})}{\tilde{\omega}^2(\mathbf{q+G})}.
\end{equation}
From Eqs.\ \eqref{eqapp:kk_plugged}, \eqref{eqapp:fsum}, and \eqref{eqapp:static}, the parameters $A(\mathbf{q+G})$, $\tilde{\omega}(\mathbf{q+G})$, and $\Omega(\mathbf{q+G})$ can be obtained.
Combining Eq.\ \eqref{eqapp:kk_plugged} and \eqref{eqapp:fsum}, we obtain $\Omega(\mathbf{q+G})$ as
\begin{equation}\label{eqapp:omegap}
    \Omega(\mathbf{q+G})=\omega_p.
\end{equation}
It can be seen that by definition, $\Omega(\mathbf{q+G})$ is a constant and is equal to $\omega_p$.
This parameter is called an effective bare plasma frequency in Ref.\ \cite{hybertsen1986electron}.
Subsequently, plugging Eq.\ \eqref{eqapp:omegap} into \eqref{eqapp:static}, we obtain
\begin{equation}\label{eqapp:omegatilde}
    \tilde{\omega}(\mathbf{q+G})=\omega_p[1-\varepsilon^{-1}(\mathbf{q+G},\omega=0)]^{-1/2},
\end{equation}
which represents the actual position of the pole of the inverse dielectric function as can be seen from Eq.\ \eqref{eqapp:ppmimag}.
Finally, plugging Eq.\ \eqref{eqapp:omegatilde} into \eqref{eqapp:fsum}, we obtain the amplitude
\begin{equation}
A(\mathbf{q+G})=-\frac{\pi}{2}\omega_p[1-\varepsilon^{-1}(\mathbf{q+G},\omega=0)]^{1/2}.
\end{equation}

With Eqs.\ \eqref{eqapp:omegap}, \eqref{eqapp:omegatilde}, and \eqref{eqapp:ppmimag}, and evaluating the integral in Eq.\ \eqref{equ:w} (considering that $\int f(x)\delta(x-a)=f(a)$), Eq.\ \eqref{eq:wbloch} is finally obtained. 
In this model, the plasma frequency $\omega_p$ is an input to obtain the dielectric function, and is computed using the $f$-sum rule with the static dielectric function from DFT using VASP. 
There are several other plasmon-pole models, that instead of enforcing the $f$-sum rule, use other constraints to obtain the frequency dependent inverse dielectric matrix\cite{hamada1990self,rohlfing1995efficient,godby1989metal}, however, these require explicit eigenvalues and eigenfunctions of the static dielectric matrix, and come with greater computational cost.
We note that  a future direction of this work could investigate more complicated PPM's, but this is not considered further in this work. 

\clearpage
\subsection{\label{app:naph_bands}Electronic structure}

\begin{figure}[!ht]
\centering
\includegraphics[width=0.5\textwidth]{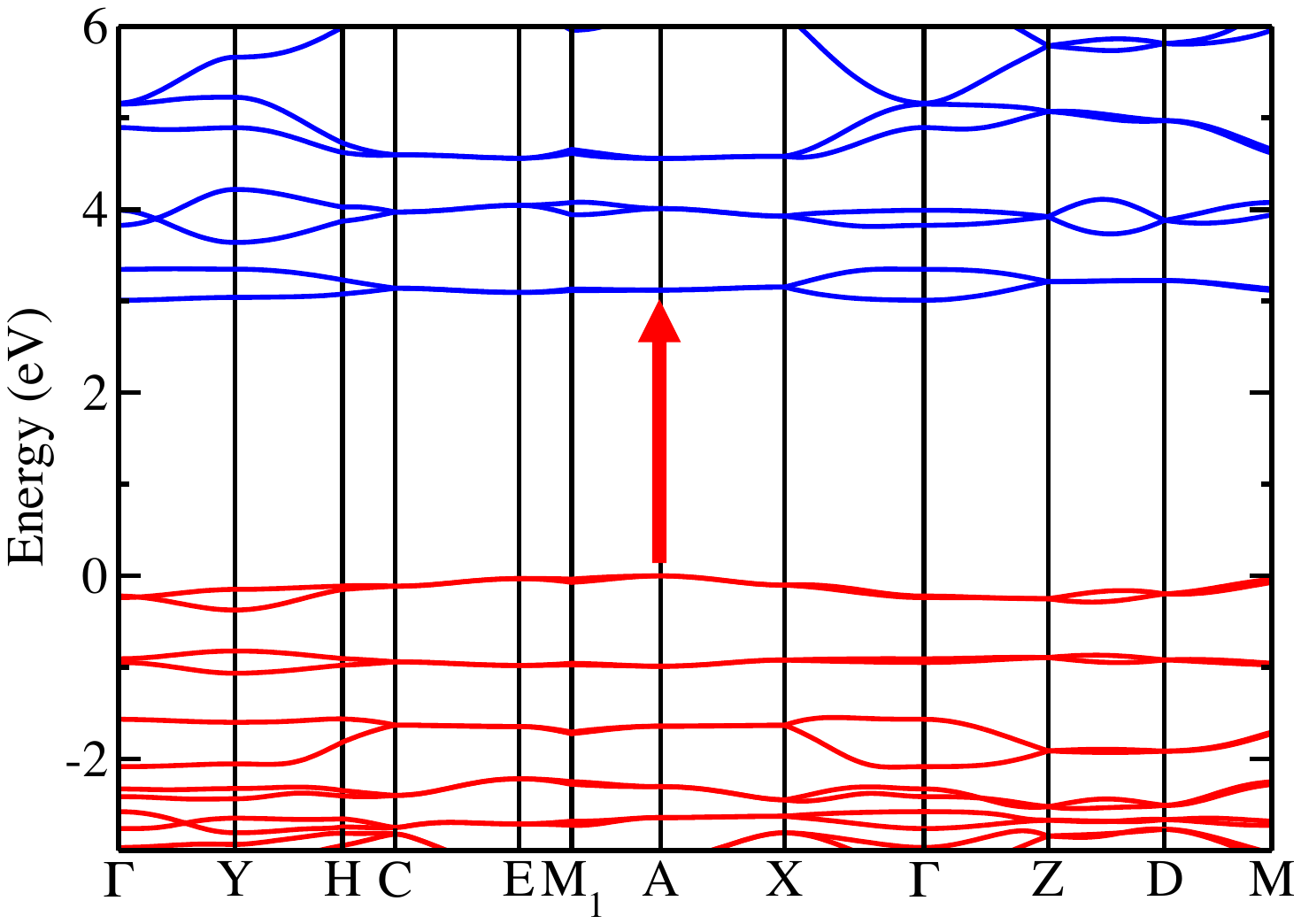}
\caption{\label{fig:app_naph_bands}
Band structure of crystalline naphthalene calculated using DFT-PBE.
The red arrow marks the lowest direct gap at the $A$ point. 
}
\end{figure}

In this section, we show the calculated electronic band structure along
a high symmetry path through the Brillouin zone generated with the AFLOW package\cite{curtarolo2012aflow}. 
While the high-symmetry path differs from that used in Ref.\ \onlinecite{hummer2005electronic},
we verified that at the same high symmetry points, good agreement is observed. 
In Fig.\ \ref{fig:app_naph_bands} we see the same feature for band gaps, as the direct gap of 3.12 eV appears at $\mathbf{A}=(0.5,0.5,0)$, and the lowest indirect gap of 3.00 eV appears between $\mathbf{A}=(0.5,0.5,0)$ and $\Gamma=(0,0,0)$. 
We note that the lattice parameters are also slightly different by less than 1.5\%.
We used the lattice parameters from a more recent experimental study\cite{capelli2006molecular}, while Ref.\ \onlinecite{hummer2005electronic} used lattice parameters from older reports\cite{cruickshank1957detailed}.
\clearpage
\subsection{\label{secapp:optics}Optical spectra with independent particle approximation}

\begin{figure}[!ht]
\centering
\includegraphics[width=0.5\textwidth]{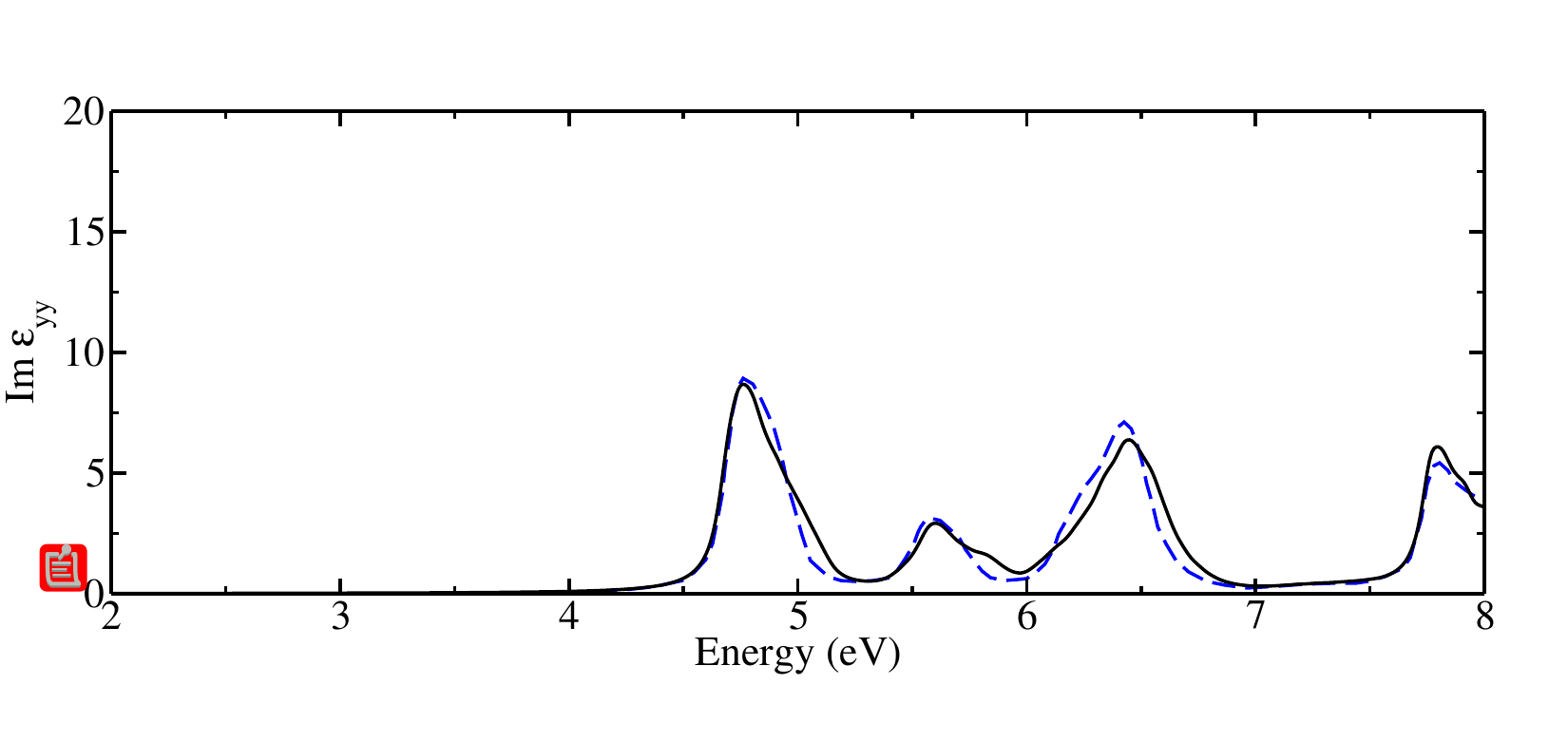}
\caption{\label{fig:sup_GGA}
Comparison between the spectra without excitonic effect from this work (black) and Ref.\ \cite{hummer2005electronic} (dashed blue).
This plot uses a life time broadening of 0.05 eV to compare to the literature result.
A scissor shift of $\Delta=1.55$ eV is applied to the spectra generated in this work. }
\end{figure}

Next, we show additional information about the calculated optical spectra without the excitonic effect. 
In the main text of the paper, we have compared our calculated optical spectra with the spectra in literature and here we provide one additional comparison of the GGA+$\Delta$ spectrum to that of Ref.\ \cite{hummer2005electronic}. 
This spectra is reported from the same authors as Ref.\ \cite{hummer2005oligoacene} but spans a wider energy range from 1 eV to 8 eV.
However, different from Ref.\ \cite{hummer2005oligoacene} and this study, a life time broadening of 0.05 eV is used instead of 0.1 eV. 
The result can be seen in Fig.\ \ref{fig:sup_GGA}.
We see that our calculated spectra agrees very well with Ref.\ \cite{hummer2005electronic} as both the peak positions and amplitudes match very well. 
In the same paper, the authors reported the calculation of the electronic band gap using DFT-GGA as 3.10 eV, and we verified that the GGA band gap of our study is 3.12 eV, which is also in excellent agreement. 

\clearpage
\subsection{Effect of the sampling}

\begin{figure}[h]
\centering
\includegraphics[width=0.5\textwidth]{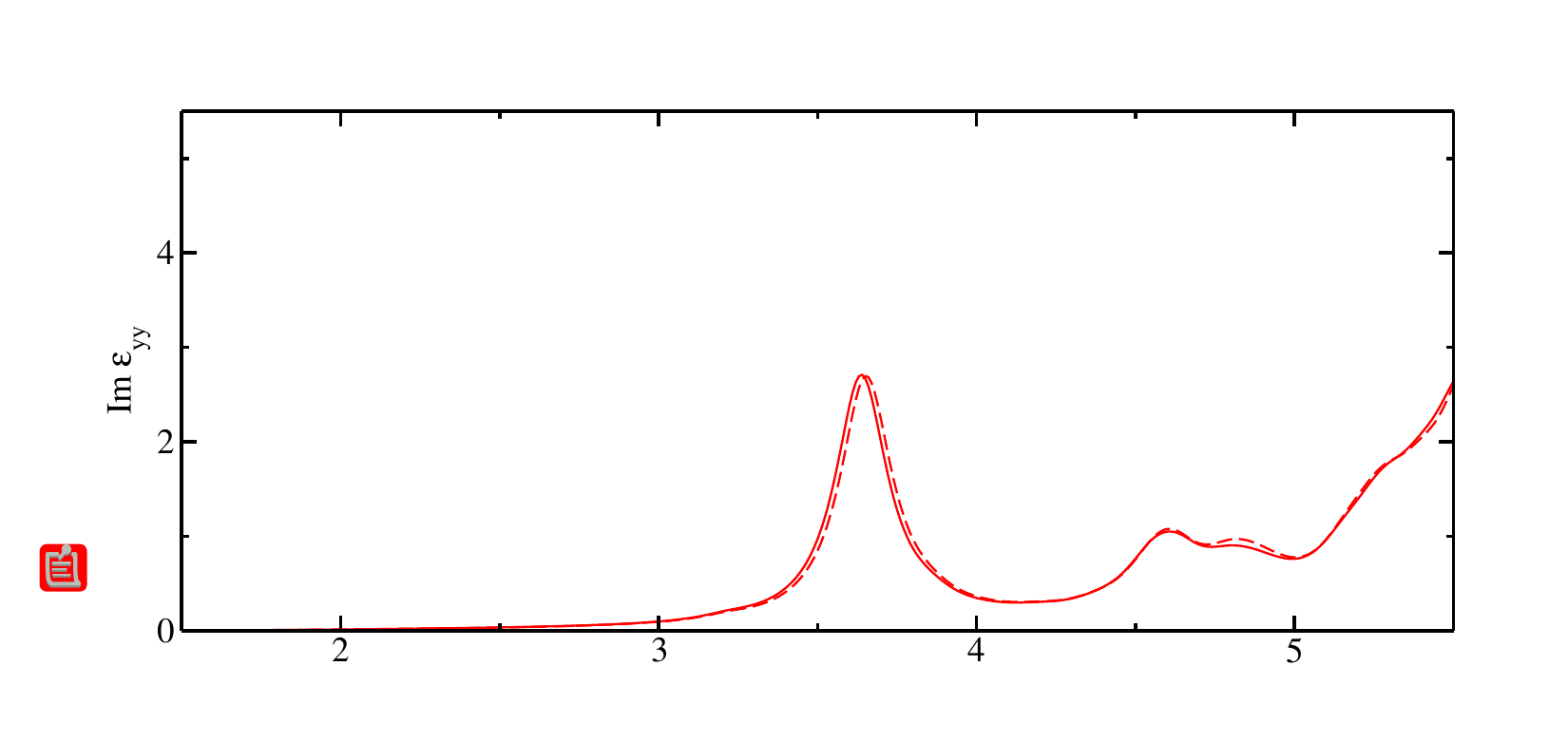}
\caption{\label{fig:sup_samp}Test for the effect of reducing the frequency sampling with red solid: 0.3 eV sampling and red dashed: 0.15 eV sampling. The spectra are calculated with a $\mathbf{k}$-point sampling of $5\times 7\times5$ and energy cutoff of 9 eV. }
\end{figure}

Here, we briefly report our test of the effect of the frequency sampling on the optical spectra with dynamical screening. 
In the main text, we adopted a frequency sampling of 0.3 eV, and we examine if we would see large effect by further reducing this to 0.15 eV. 
The comparison is shown in Fig.\ \ref{fig:sup_samp}.
It can be seen from the figure that the spectrum is not affected significantly by the sampling of the frequency grid. 
Slight difference can be seen between the two curves, as the position of the first peak is changed by less than 0.01 eV. 
However, the overall shape and the peaks are well described by a sampling of 0.3 eV, which suggests that further sampling of the frequency grid is not necessary. 
As a denser sampling of the frequency grid can be very expensive, due to that it requires to solve the BSE eigenvalue problem at many more frequencies, we used 0.3 eV across all dynamical simulations in the main text. 

\clearpage
\subsection{Convergence of spectra}

\begin{figure}
\centering
\includegraphics[width=0.5\textwidth]{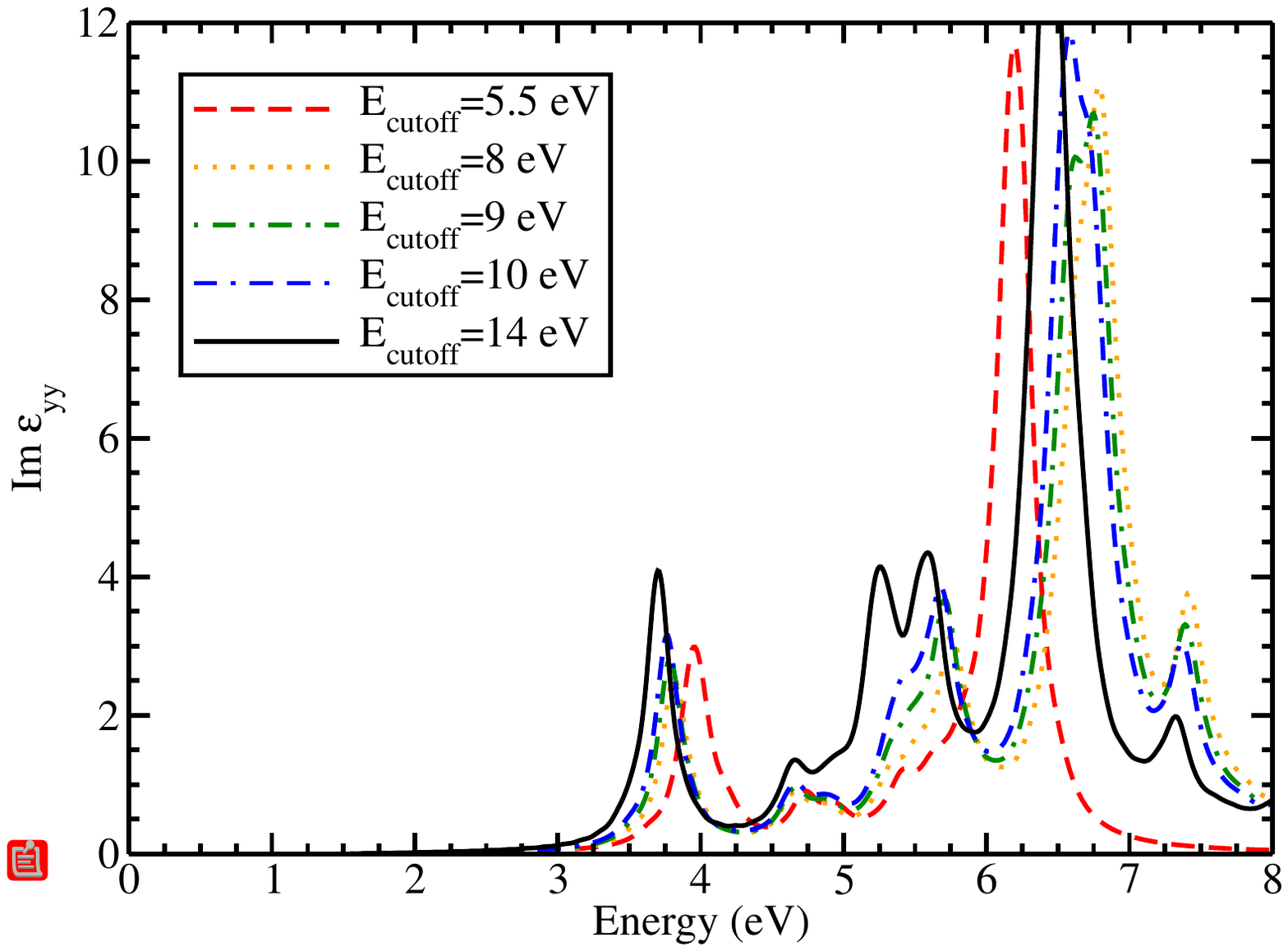}
\caption{\label{fig:conv_spectra}
(Color online.) Calculated imaginary part of the dielectric function for light polarization $\parallel b$ direction for various BSE energy cutoffs. 
The value of the cutoffs are before a scissor shift of 1.55 eV is applied. 
The lowest peaks are well-predicted by an energy cutoff of 9 eV, while the peaks after 5 eV are underconverged. 
}
\end{figure}

We tested the convergence of the spectra with respect to the BSE energy cutoff within the static approximation. 
Figure \ref{fig:conv_spectra} shows the calculated imaginary part of the dielectric function for light polarization $\parallel b$ direction using an $A$-centered $\mathbf{k}$-point grid of $5\times7\times5$ with various BSE energy cutoffs. 
In the main text, we used an energy cutoff of 14 eV to compare to spectra in the literature, while for investigating dynamical effects, we reduced the cutoff to 9 eV due to the increased computational cost. 
It can be seen from the figure that the peaks above 5 eV are underconverged, thus for investigating dynamical effects, we focus on the peaks below 5 eV. 

\clearpage
\subsection{\label{sec:app_conv_ex}Convergence of the static exciton-binding energy}

We tested for the convergence of the exciton-binding energy with respect to two parameters, within the static approximation of the screening:
the $\mathbf{k}$-point sampling and energy cutoff of the BSE matrix.
The test with respect to energy cutoff is performed up to 14 eV. 
The $\mathbf{k}$-point samplings tested are listed in Table \ref{tab:kptconv}. 
In our simulation, a hybrid $\mathbf{k}$-point sampling is adopted so that the part close to the $A$ point in the Brillouin zone, where the lowest vertical transition is located, is well sampled, while the outer part is sampled with a relatively coarse grid. 
In Table \ref{tab:kptconv}, we list the sampling of the outer part of the BZ, the fraction of the inner part, the sampling of the inner part, and the effective sampling of the BZ.
The details of this notation can be found in Ref.\ \cite{fuchs2008efficient}. 
We adopted an energy cutoff of 5.5 eV for the $\mathbf{k}$-point sampling tests. 
The densest sampling contains over 5500 $\mathbf{k}$-points, and the corresponding BSE matrix size is over 140,000.

\begin{figure}
\centering
\includegraphics[width=0.5\textwidth]{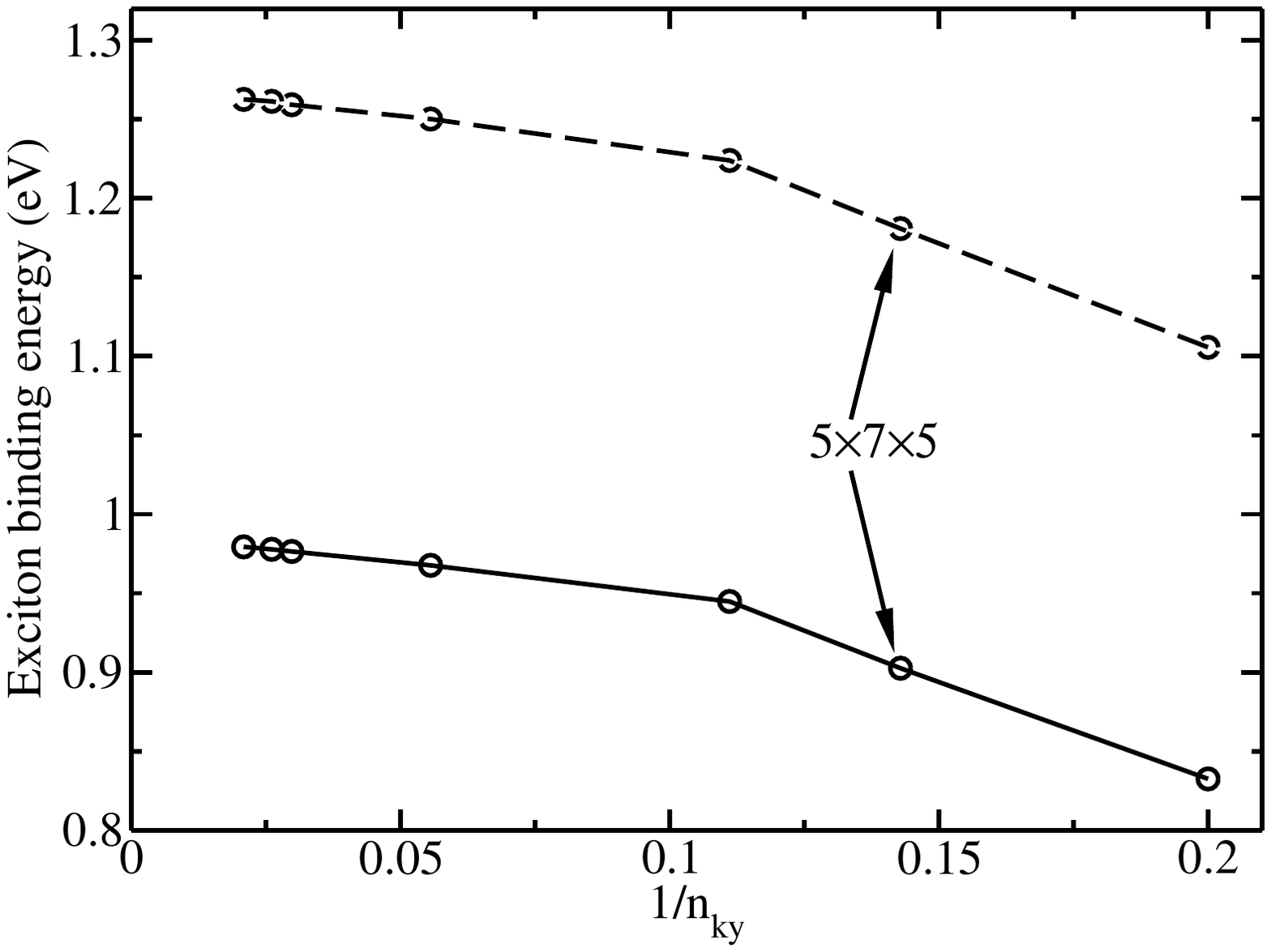}
\caption{\label{fig:conv_kpt}
Convergence of the exciton-binding energy of with adopting the static approximation.
A $\mathbf{k}$-point sampling of $5\times7\times5$ is used to examine the dynamical screening effect, corresponding to the second last point to the right.
Both the exciton binding of the lowest dark states (dashed) and the first major peak, which is what typically is referred to in the literature (solid), are shown. }
\end{figure}

\begin{figure}
\centering
\includegraphics[width=0.45\textwidth]{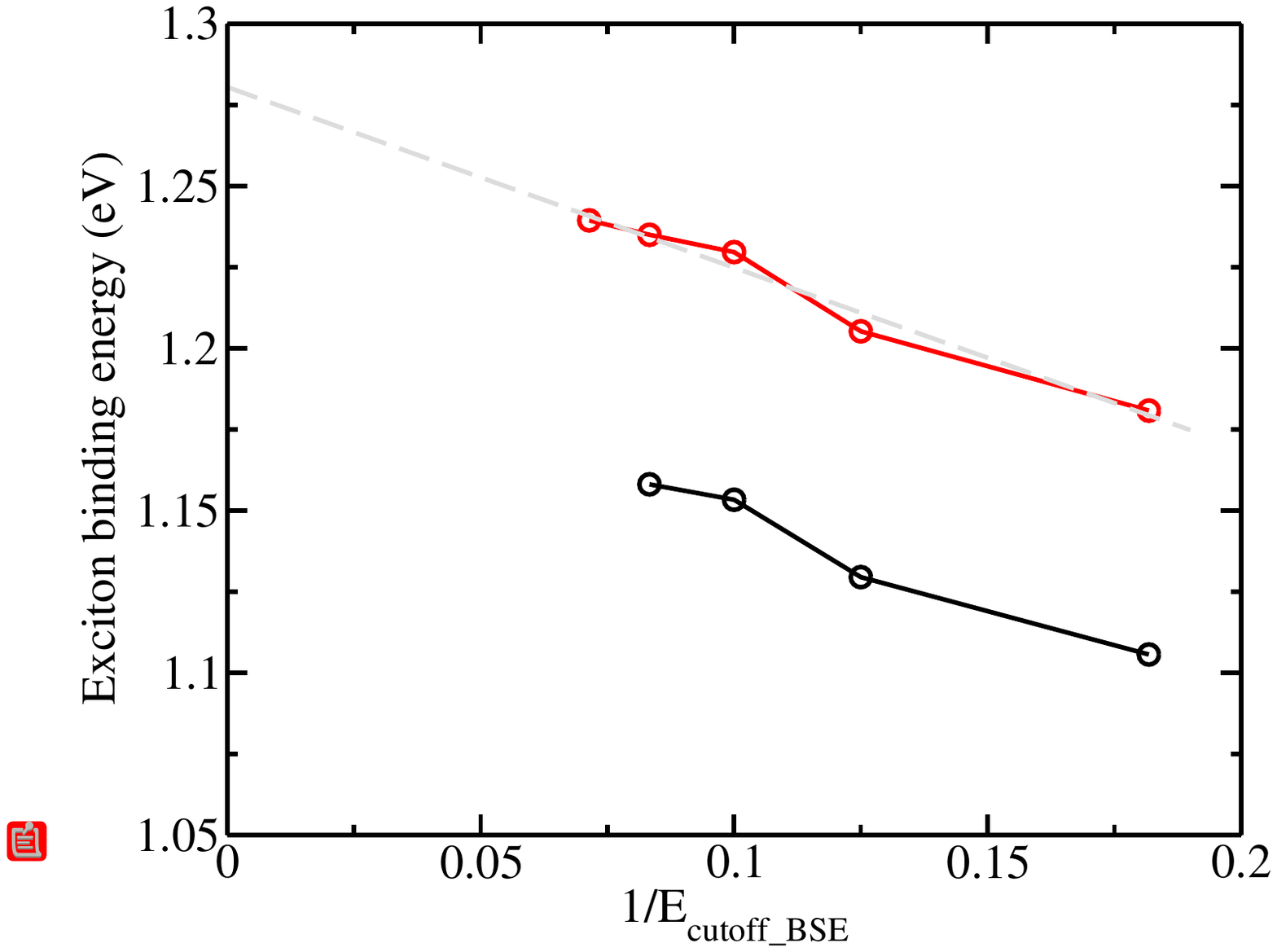}
\includegraphics[width=0.45\textwidth]{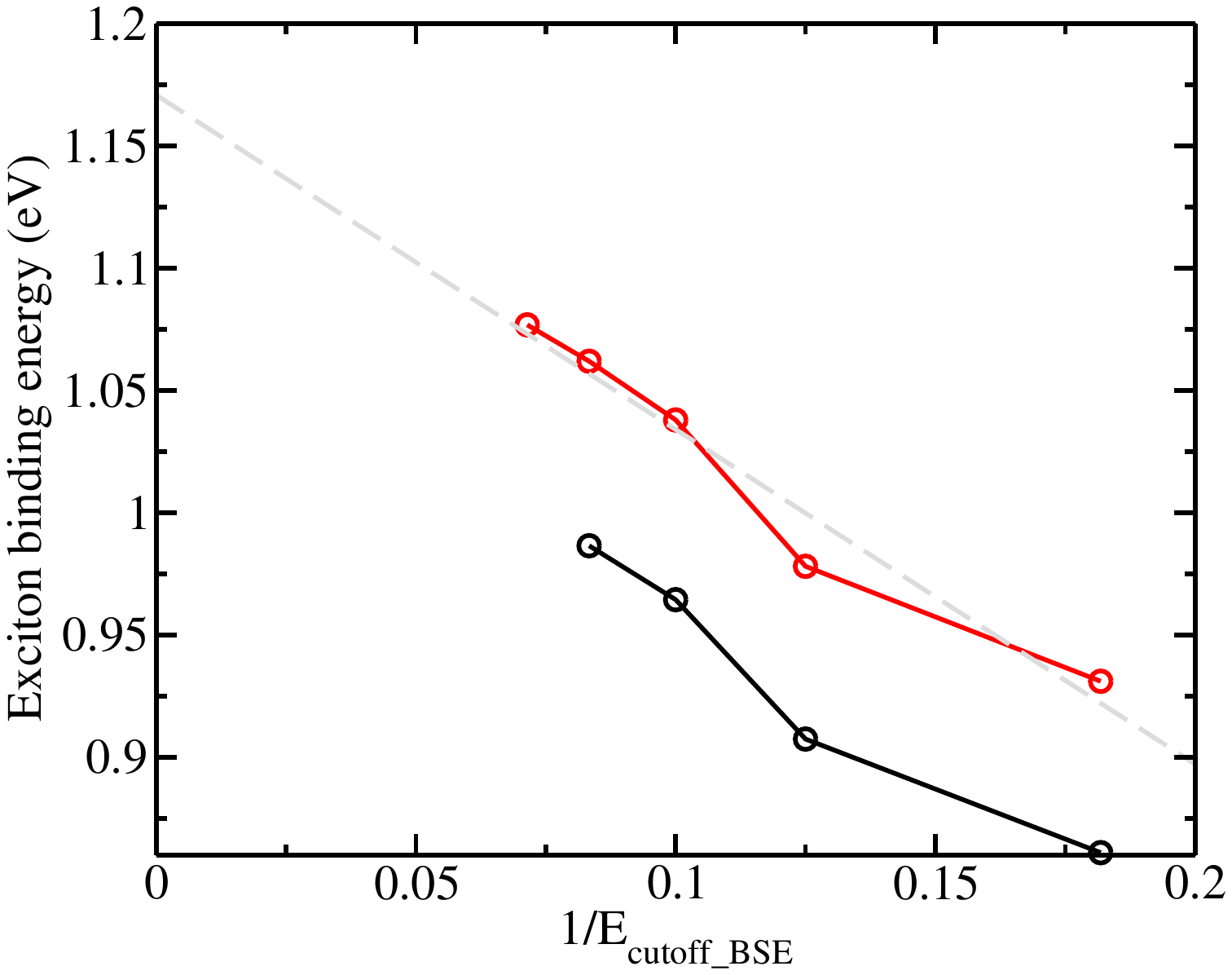}
\caption{\label{fig:conv_ecutoff}
Convergence of the exciton binding energy from static approximation versus the BSE energy cutoff.
The two graphs are exciton binding energies for the lowest dark excitonic state (left) and for the first major peak (right).
The two curves on each graph are for different $\mathbf{k}$-point samplings with black: $3\times5\times3$ and red: $5\times7\times5$.}
\end{figure}

The result of the convergence tests of the exciton-binding energy with respect to $\mathbf{k}$-point sampling and energy cutoff can be found in Fig.\ \ref{fig:conv_kpt} and Fig.\ \ref{fig:conv_ecutoff}. 
In the naphthalene crystal, we see that the lowest eigenvalues of the BSE Hamiltonian are dark excitons with very low oscillator strength. 
We note that in the literature, exciton binding is reported for the excitonic states of the first major peak in the y-polarization\cite{hummer2005oligoacene}.
In our test for the convergence of the static eigenvalues with respect to $\mathbf{k}$-points and energy cutoff, we tested for both the lowest eigenvalue and the first major peak. 

\begin{table}
\centering
\caption{\label{tab:kptconv}
The $\mathbf{k}$-point grids used for convergence tests within the static approximation, following the notation of Ref.\ \cite{fuchs2008efficient}.}
    \begin{tabular}{cccc}
    \hline
         Outer samp.&Inner frac.&Inner samp.&Effective samp.\\ \hline
         $3\times5\times3$&$3\times5\times3$&$3\times5\times3$&$3\times5\times3$  \\
         $5\times7\times5$& $5\times7\times5$&$5\times7\times5$&$5\times7\times5$ \\
         $7\times9\times7$&$7\times9\times7$&$7\times9\times7$&$7\times9\times7$\\
         $7\times9\times7$&$4\times6\times4$&$9\times13\times9$&$14\times18\times14$\\
         $8\times12\times8$&$3\times5\times3$&$11\times15\times11$&$26.7\times33.6\times26.7$\\
         $8\times12\times8$&$3\times5\times3$&$13\times19\times13$&$32\times43.2\times32$\\
         $8\times12\times8$&$3\times5\times3$&$15\times21\times15$&$37.3\times48\times37.3$\\
         \hline
    \end{tabular}
    
    \label{tab:my_label}
\end{table}
We used a combination of $5\times7\times5$ $\mathbf{k}$-points and 9 eV energy cutoff to examine dynamical screening effects. 
From the plots, interpolating to zero allows to estimate the converged value at infinite $\mathbf{k}$-point sampling and BSE energy cutoff. 
We find that the convergence of the energy cutoff is more complicated since the two curves for the dark and bright excitonic state differ much more than in the case of $\mathbf{k}$-point convergence. 
However, linear interpolation can still be performed when excluding the smallest energy cutoff of $E_{\text{cutoff}}=5.5$ eV. 
For the $\mathbf{k}$-point sampling test, we estimate that compared to the extrapolated value of 0.99 eV, the choice of $5\times7\times5$ induces an underestimation of 0.06 eV for the exciton-binding energy. 
Further, for the energy cutoff test, the choice of the energy cutoff of 9 eV induces an additional underestimation of 0.05 eV for the dark state and 0.17 eV for the major peak compare to the extrapolated value of 1.28 eV and 1.17 eV, respectively.
In combination, for the lowest bright excitonic state, we estimate a total underestimation of 0.23 eV for the exciton binding energy due to the choice of $\mathbf{k}$-points sampling and energy cutoff.
We note that although the underestimation is not negligible, it is not related to actual physical phenomena of dynamical screening that we intend to study in the system, and thus does not cause fundamental differences in the results presented in the main text.

\end{document}